\documentclass[preprint,showpacs,showkeys,aps,prd,amsfonts,eqsecnum,nofootinbib]
{revtex4}
\usepackage{graphicx,epsfig}
       {\left[\begin{array}{>{\displaystyle}c}}%
       {\end{array}\right]}
       {\left\{\begin{array}{l}}%
       {\end{array}\right\}}

\begin{document}
\preprint{CUMQ/HEP 139}
%
%
\title{Single Top Production via Gluon Fusion at CERN LHC}
\author{Gad Eilam}\email[]{eilam@physics.technion.ac.il}
\affiliation{Technion-Israel Institute of Technology, 32000 Haifa, ISRAEL}
\author{Mariana Frank}\email[]{mfrank@vax2.concordia.ca}
\author{Ismail Turan}\email[]{ituran@physics.concordia.ca}
\affiliation{Department of Physics, Concordia University, 7141
Sherbrooke Street West, Montreal, Quebec, CANADA H4B 1R6}
\date{\today}

\begin{abstract}
We calculate the one-loop flavor violating top quark decay
$t \to cgg$ in the  Minimal Supersymmetric Standard Model.  We
discuss the branching ratios obtained with minimal flavor violation,
as well as with soft-supersymmetry induced general flavor violation.
Based on this rate we calculate the cross section for the single top
quark production via gluon fusion, $gg \to t {\bar c}$, and evaluate
its contribution to the cross section for single top quark production
in $pp$ collisions at the Large Hadron Collider.
We calculate all contributions coming from the
standard model and  charged Higgs loops, as well as gluino (and
neutralino)-up-type squarks, and chargino-down-type squarks loops.
Our numerical results show that the gluino and the chargino contributions
are largest over the whole parameter range in
the unconstrained Minimal Supersymmetric Standard Model.
While in general the gluino contributions
dominate the cross section, this result depends on the supersymmetric
flavor violating parameters in the up and down squark sector, the
relative mass of the
gauginos, and whether or not the Grand Unified Theory relationships
between gaugino masses
are satisfied. In the
most promising scenarios, the $pp\to t\bar{c}+\bar{t}c +X $ cross section
at the Large Hadron Collider can
reach a few hundreds fb.
\pacs{12.60.Jv, 11.30.Hv, 14.65.Ha}
\keywords{Rare Top Decays, Single Top Production, MSSM, Higher-Order Dominance}
\end{abstract}
\maketitle
\section{Introduction}\label{sec:intro}
One of the main goals at the CERN Large Hadron Collider (LHC) is to
study the production and decay of top quarks. The importance
of studying the physics of the top is obvious. It is the quark
which is closest to the scale
of electroweak symmetry breaking and is therefore most sensitive
to that scale, and thus to New Physics (NP) beyond the Standard Model (SM).
One of the important tests of the SM
is its predictions for the yield of single tops in
hadronic collisions. The measurement
of single top production cross sections has turned out to be a
challenging task so far \cite{Taffard:2005rk} and only upper
limits are obtained. For instance, the D0 experiment, at Tevatron
II with integrated luminosity of $230 {\rm fb}^{-1}$, obtained
the following upper limits on the $s~(t)$-channel processes
(as defined below):
6.4 (5.0) pb, at $95\%$ C.L.
It is expected that increased luminosity and
improved methods of analysis will eventually lead to the detection of
single top events in Tevatron II and subsequently at the LHC.

Single top production in hadronic machines has been thoroughly
discussed within the SM where, at lowest order,
one has the tree level contributions of
$s$-channel ($q{\bar q}\to t{\bar b}$ through $W$ exchange),
$t$-channel ($u b\to t d$ via $W$ exchange) and $g b\to tW$
with a top quark exchanged. In \cite{Sullivan:2004ie,Cao:2004ap}
one finds the
most recent SM results, which include
Next to Leading Order (NLO) corrections. These are predicted  to be
approximately equal to (all
the following cross sections are in pb),
6.6 (4.1) for a single
$t$ (${\bar t}$) production in the
$s$-channel, and 156 (91) for a single $t$ (${\bar t}$) production
in the $t$-channel at LHC \cite{Sullivan:2004ie}.
The background for single top production in the
SM was estimated in \cite{Sullivan:2005ar}.

At the same time, there has been an increased
interest in studying forbidden or highly suppressed processes as they
appear ideal for finding the physics lying beyond the SM.
As alluded to before,
top quark interactions, in particular, might provide a fertile ground
to searches for NP. It is expected that if NP
is associated with the mass generation mechanism, it may be
more apparent in top quark interactions, rather than in the light
fermion sector. Along these lines, there have been suggestions that the
Flavor Changing Neutral Currents (FCNC) single top quark production
could be rather sensitive to non-SM
couplings such as $tcV~(V=g,~\gamma,~Z)$ and $tcH$ \cite{Han:1995pk}.
The advantage in looking for
FCNC processes in top physics is that although these
exist in the SM, they are minute,
leading to tiny, unmeasurable SM effects. In general, any
measurable FCNC process involving the top will indicate that one is witnessing
the effects of NP.  Note that here we are only interested in
processes that are driven by FCNC couplings, which are highly suppressed
in the SM by the GIM mechanism. Therefore we do not consider NP
corrections to SM couplings (like $tbW$ or $Zqq$)
or the contributions of new particles (either external or internal),
like $Z^\prime$ or $W^\prime$,
except those of the supersymmetric partners of SM particles.

FCNC effects in top production contribute to the following single top
production processes on the partonic level:
$cg\to t$,
$cg\to tg$,
$cq({\bar q})\to tq({\bar q})$,
$q{\bar q}\to t{\bar c}$ and
$gg\to t{\bar c}$,
as well as all the above with $c\longrightarrow u$.
These subprocesses have been investigated
in the presence of FCNC effective couplings and in the framework of various
NP models \cite{Han:1995pk}.

Of all scenarios of physics beyond the SM, supersymmetry
is the most popular. A characteristic feature of supersymmetry is
that, in addition to the SM FCNC generated by the
Cabibbo-Kobayashi-Maskawa (CKM) mixing matrix, it can provide large
soft supersymmetry-generated FCNC which would enhance rates and cross
sections beyond SM values.  The proton collider LHC can produce
supersymmetric particles, such as squarks and gluinos, with masses up
to 3 TeV; as well as potentially lighter ones, such as
charginos/neutralinos. Flavor-changing interactions appear in
supersymmetry in loops involving these particles, and thus
enhancements in FCNC signals are expected at the LHC.

Single top quark production generated through FCNC processes has
been discussed within the effective Lagrangian formalism  in a model
independent way \cite{Malkawi:1995dm}, as well as in model-dependent
scenarios \cite{Li:1999ms}. The purpose of this study is to analyze
one such class of rare single quark FCNC production: the gluon fusion
$gg \to t {\bar c}$ within the framework of low-energy supersymmetry.
This process was analyzed in \cite{Liu:2004bb} where QCD-only loops
(loops of gluino and squarks), were evaluated in the context of the
unconstrained Minimal Supersymmetric Standard Model (MSSM).
However it is known from analyses of $t\to cV$
that charginos, and sometimes neutralinos, can have a large effect on
FCNC.  Here we  first discuss the rare decay $t \to cgg$ and show it
to be larger than $t \to cg $ over most of the parameter space in
certain cases. Then
we perform a complete analysis of  $gg \to t {\bar c}$ in both the
constrained MSSM
(where FCNC decays and cross sections are driven by
chargino-down-like squark loops) and the unconstrained MSSM (where
gluino and neutralino loops contribute as well). We include the SM
and charged Higgs contributions, contributions from
chargino, neutralinos and gluino loops, as well as interference
effects between SM and non-SM effects, in the context of
the most general left-left, left-right and right-right
intergenerational squark mixings.  We also address the
observability of these channels at LHC.

Our paper is organized as follows: After a description of the FCNC
sources in the unconstrained MSSM (Section II),  we present  our
complete analysis of the branching ratio for the top quark $t \to
cgg$ in MSSM, and compare it to the SM case, where $t\to cgg$ was
shown to be larger than $t \to cg$ \cite{EFTSM} (Section III).
Section IV is devoted to the calculation of the gluon fusion cross
section $gg \to t{\bar c}$, as well as the
    evaluation of the cross section for $pp \to t{\bar c}+X$ at the LHC
through gluon fusion. We include a detailed numerical analysis of the
various relative supersymmetric contributions from gluino and
chargino loops with or without Grand Unified Theory (GUT) mass relations,
in addition to a comparison
of the constrained versus the unconstrained MSSM predictions, as well as
  observability of these channels. Our
conclusions and prospects for experimental observations are presented
in Section V.

\section{FCNC in the unconstrained MSSM}\label{sec:fcnc}
In the unconstrained MSSM there are two sources of flavor
violation.  The first one arises from the  different mixing of quarks
in the $d$- and $u$-sectors
in the physical bases, and it is described by the CKM matrix
(inherited from the SM).  In the minimal version of MSSM (the
constrained MSSM) this is the only source of flavor violation.
The second source of flavor violation consists of a
possible misalignment between the rotations that
diagonalize the quark and squark sectors, and it is a characteristic of
soft supersymmetry breaking. We  work in the most general version of
the model and discuss the constrained version as a limit.
The superpotential of the MSSM Lagrangian is
\begin{equation}
      \label{eq:W} {\mathcal{W}} = \mu H^1 H^2 + Y_l^{ij} H^1
{L}^i {e}_R^j + Y_d^{ij} H^1 {Q}^i {d}_R^j
+ Y_u^{ij} H^2 {Q}^i {u}_R^j.
\label{eq:superpot}
\end{equation}

The part of the soft-SUSY-breaking Lagrangian responsible for
the non-minimal squark family mixing is given by
\begin{eqnarray}
\label{eq:lagrangian}
\mathcal{L}^{\text{squark}}_{\text{soft}} &=&
-\tilde Q^{i\dagger} (M_{\tilde Q}^2)_{ij} \tilde Q^j
-\tilde u^{i\dagger} (M_{\tilde U}^2)_{ij} \tilde u^j
-\tilde d^{i\dagger} (M_{\tilde D}^2)_{ij} \tilde d^j \nonumber \\
&&\quad + Y_u^i A_u^{ij} \tilde Q_i H^2 \tilde u_j
+ Y_d^i A_d^{ij} \tilde Q_i H^1 \tilde d_j\,.
\end{eqnarray}
In the above expressions $ Q$ is the SU(2) scalar doublet, $ u$, $ d$ are
the up- and down-quark SU(2) singlets ($\tilde Q, \tilde u, \tilde d$
represent scalar quarks), respectively, $Y_{u,d}$ are the
Yukawa couplings and $i,j$ are generation indices.  The flavor-changing
effects come from the non-diagonal entries in the bilinear terms
$M_{\tilde Q}^2$, $M_{\tilde U}^2$, and $M_{\tilde D}^2$, and from the
trilinear terms $A_u$ and $A_d$. Here $H^{1,2}$ represent two $SU(2)$
Higgs doublets with vacuum expectation values
\begin{equation}
      \langle H^1 \rangle = \left( \begin{array}{c}
\frac{v_1}{\sqrt{2}} \\ 0 \end{array} \right) \equiv \left(
\begin{array}{c} \frac{v \cos \beta}{\sqrt{2}} \\ 0 \end{array}
\right), \hspace{2cm} \langle H^2 \rangle = \left(
\begin{array}{c} 0 \\ \frac{v_2}{\sqrt{2}} \end{array} \right) \equiv
\left( \begin{array}{c} 0 \\ \frac{v \sin \beta}{\sqrt{2}}
\end{array} \right),
\end{equation}
where $v=(\sqrt{2}G_F)^{-1/2}=246$~GeV, and the angle $\beta$ is
defined by $\tan \beta\equiv v_2/v_1$, the ratio of the vacuum
expectation values of the two Higgs doublets and $\mu$ is the Higgs
mixing parameter.

Since we are concerned with top quark physics, we assume that the
non-CKM squark mixing is significant only for
transitions between the
squarks of the second and third generations.  These mixings are expected to be
the largest in Grand Unified Models  and are also experimentally the
least constrained.  The most stringent bounds on these transitions
come from $b \to s \gamma $.  In
contrast, there exist strong experimental bounds involving the first
squark generation, based on data from $K^0$--$\bar K^0$ and $D^0$--$\bar
D^0$ mixing~\cite{Gabbiani:1996hi}.

It is convenient to specify the squark mass matrices in the
super-CKM basis, in which the mass matrices of the quark fields
are diagonalized by rotating the superfields.
Our parameterization of the flavor-non-diagonal squark mass
matrices for the up- and down-type squarks, for the MSSM with real
parameters, reads as follows,
\begin{equation}
\label{eq:usquarkmass}
M^2_{\tilde u} =
\left( \begin{array}{cccccc}
M_{{\tilde L} u}^2 & 0 & 0 & m_u {\cal A}_u & 0 & 0  \\
0 & M_{{\tilde L} c}^2 & (M^2_{\tilde{U}})_{LL} & 0 & m_c {\cal A}_c
&(M^2_{\tilde{U}})_{LR} \\
0 & (M^2_{\tilde{U}})_{LL} & M_{{\tilde L} t}^2 & 0 &
(M^2_{\tilde{U}})_{RL} & m_t {\cal A}_t \\[.3ex]
m_u {\cal A}_u & 0 & 0 & M_{{\tilde R} u}^2 & 0 & 0 \\
0 & m_c {\cal A}_c & (M^2_{\tilde{U}})_{RL} & 0 &M_{{\tilde R} c}^2 &
(M^2_{\tilde{U}})_{RR} \\
0 & (M^2_{\tilde{U}})_{LR} & m_t {\cal A}_t & 0 &
(M^2_{\tilde{U}})_{RR} &M_{{\tilde R} t}^2
\end{array}\right)\,,
\end{equation}
\begin{equation}
\label{eq:dsquarkmass}
M^2_{\tilde d} = \left(\begin{array}{cccccc}
M_{{\tilde L} d}^2 & 0 & 0 & m_d {\cal A}_d & 0 & 0 \\
0 & M_{{\tilde L} s}^2 & (M^2_{\tilde{D}})_{LL} & 0 & m_s {\cal A}_s
&(M^2_{\tilde{D}})_{LR} \\
0 & (M^2_{\tilde{D}})_{LL} & M_{{\tilde L} b}^2 & 0 &
(M^2_{\tilde{D}})_{RL} & m_b {\cal A}_b \\[.3ex]
m_d {\cal A}_d & 0 & 0 & M_{{\tilde R} d}^2 & 0 & 0 \\
0 & m_s {\cal A}_s &(M^2_{\tilde{D}})_{RL} & 0 &M_{{\tilde R} s}^2 &
(M^2_{\tilde{D}})_{RR} \\
0 & (M^2_{\tilde{D}})_{LR} & m_b {\cal A}_b & 0 &
(M^2_{\tilde{D}})_{RR} &M_{{\tilde R} b}^2
\end{array}\right)\,,
\end{equation}
where
\begin{eqnarray}
\label{eq:squarkparam}
M_{{\tilde L}q}^2 &=&
       M_{\tilde Q,q}^2 + m_q^2 + \cos2\beta (T_q - Q_q s_W^2) m_Z^2\,,
\nonumber \\
M_{{\tilde R}\{u,c,t\}}^2 &=&
       M_{\tilde U,\{u,c,t\}}^2 + m_{u,c,t}^2 + \cos2\beta Q_t s_W^2
m_Z^2\,, \nonumber \\
M_{{\tilde R}\{d,s,b\}}^2 &=&
       M_{\tilde D,\{d,s,b\}}^2 + m_{d,s,b}^2 + \cos2\beta Q_b s_W^2 m_Z^2\,, \\
{\cal A}_{u,c,t} &=& A_{u,c,t} - \mu\cot\beta\,, \nonumber \\
{\cal A}_{d,s,b} &= &A_{d,s,b} - \mu\tan\beta\,, \nonumber
\end{eqnarray}
with $m_q$, $T_q$, $Q_q$ the mass, isospin, and electric charge of the
quark $q$, $m_Z$ the $Z$-boson mass, $s_W \equiv \sin\theta_W$ and
$\theta_W$ the electroweak mixing angle. In the above matrices we
assumed that significant mixing occurs between the second and third
generations only.

We define the dimensionless flavor-changing parameters
$(\delta_{U,D}^{23})_{AB}$ $(AB = LL,\,LR,\,RL,\,RR)$ from the
flavor off-diagonal elements of the squark mass matrices Eqs.~
(\ref{eq:usquarkmass}) and (\ref{eq:dsquarkmass}) in the following
way. To simplify the calculation we assume
that all diagonal entries in $(M^2_{\tilde{U}})_{LL}$,
$(M^2_{\tilde{U}})_{LR}$, $(M^2_{\tilde{U}})_{RL}$ and
$(M^2_{\tilde{U}})_{RR}$ and similarly for
$(M^2_{\tilde{D}})_{AB}$, are set equal to the common value
$M^2_{\rm{SUSY}}$, and then we normalize the off-diagonal elements
to $M^2_{\rm{SUSY}}$  \cite{Harnik:2002vs,Besmer:2001cj},
\begin{eqnarray}
&& (\delta_{U}^{ij})_{LL} =
\frac{(M^2_{\tilde{U}})_{LL}^{ij}}{M^2_{\rm{SUSY}}}\,,\hspace{1.0truecm}(\delta_{D}^{ij})_{LL}
=
\frac{(M^2_{\tilde{D}})_{LL}^{ij}}{M^2_{\rm{SUSY}}} \nonumber \\
&& (\delta_{U}^{ij})_{RR} =
\frac{(M^2_{\tilde{U}})_{RR}^{ij}}{M^2_{\rm{SUSY}}}\,,
\hspace{1.0truecm} (\delta_{D}^{ij})_{RR}
=\frac{(M^2_{\tilde{D}})_{RR}^{ij}}{M^2_{\rm{SUSY}}}\nonumber \\
&& (\delta_{U}^{ij})_{LR} =
\frac{(M^2_{\tilde{U}})_{LR}^{ij}}{M^2_{\rm{SUSY}}}\,,\hspace{1.0truecm}(\delta_{D}^{ij})_{LR}
=
\frac{(M^2_{\tilde{D}})_{LR}^{ij}}{M^2_{\rm{SUSY}}} \nonumber \\
&& (\delta_{U}^{ij})_{RL} =
\frac{(M^2_{\tilde{U}})_{RL}^{ij}}{M^2_{\rm{SUSY}}}\,,
\hspace{1.0truecm} (\delta_{D}^{ij})_{RL}
=\frac{(M^2_{\tilde{D}})_{RL}^{ij}}{M^2_{\rm{SUSY}}}
\hspace{1.0truecm} (i \ne j,\;i,j=2,3). \label{deltadefb}
\end{eqnarray}
      The matrix ${\cal M}^2_{\tilde{u}}$ can
further be diagonalized by an additional $6\times 6$ unitary
matrix $\Gamma_U$ to give the up squark mass eigenvalues
\begin{eqnarray}
\left({\cal M}^2_{\tilde{u}}\right)^{\text{diag}} = \Gamma_U^{\dagger}
{\cal M}^2_{\tilde{u}} \Gamma_U
      \label{eq:gammaudef}.
\end{eqnarray}
For the down squark mass matrix, we also can define
${\mathcal{M}}_{\tilde{d}}^2$ as the similar form of
Eq.~(\ref{eq:gammaudef}) with the replacement of
$(M^2_{\tilde{U}})_{AB}$ ($A,B=L,R$) by $(M^2_{\tilde{D}})_{AB}$.
Note that while $SU(2)_L$ gauge invariance implies that
$(M^2_{\tilde{U}})_{LL} = K_{CKM} (M^2_{\tilde{D}})_{LL} K_{CKM}^\dagger$, the
matrices $(M^2_{\tilde{U}})_{LL}$ and $(M^2_{\tilde{D}})_{LL}$ are
correlated. Since the connecting equations are rather complicated and
contain several unknown parameters, we proceed by including the
flavor changing parameters $(\delta_{U,D}^{ij})_{AB} $ as independent
quantities, while restricting them using previously set bounds
\cite{Gabbiani:1996hi}.

Thus, in the super-CKM basis, there are potentially new sources of
flavor-changing neutral currents: Chargino-quark-squark couplings,
neutralino-quark-squark coupling
and gluino-quark-squark coupling, which arise from the
off-diagonal elements of $(M^2_{{\tilde U}, {\tilde D}})_{LL}$,
$(M^2_{{\tilde U},{\tilde D}})_{LR}$ and $(M^2_{{\tilde U}, {\tilde
D}})_{RR}$. Previous
considerations of flavor violating decays \cite{Lopez:1997xv}
      in the MSSM have shown that both up and down squarks contribute
significantly. Our analysis shows that this is the case here too, and
which one is dominant depends on the parameters of the model, and in
particular on the relative mass hierarchy between the chargino and
the gluino.

In the super-CKM basis, the quark-up squark-gluino ($\tilde g$)
interaction is given by
\begin{equation}
\mathcal{L}_{u \tilde{u} \tilde g}= \sum_{i=1}^{3}\sqrt{2}\, g_s \,
T^r_{st} \left[ \bar
u^{s}_i \,(\Gamma_U)^{ia}\,P_L\, \tilde g^r \,\tilde u^{t}_a - \bar
u^{s}_i \,(\Gamma_U)^{(i+3)a}\,P_R \,\tilde g^r \,\tilde u^{t}_a +
\text{H.c.} \right]\,,
\end{equation}
where $T^{r}$ are the $SU(3)_{c}$ generators,  $P_{L,R}\equiv (1\mp
\gamma_5)/2$, $i=1,2,3$ is the generation index, $a=1, \ldots, 6$ is
the scalar quark index, and $s,t$ are color indices. In the gluino
interaction, the flavor changing effects from soft broken
terms $M^2_{\tilde Q}$, $M^2_{\tilde U}$ and $A_{u}$ on the observables are
introduced through the matrix $\Gamma_U$.

The relevant Lagrangian terms for the
      quark-down squark-chargino ($\tilde {\chi}^{\pm}_\sigma$)
interaction are given by
\begin{eqnarray}
\mathcal{L}_{u\tilde{d}\tilde{\chi}^{+}}\!\!\!&=\!\!\!&\sum_{\sigma=1}^{2}\,
\sum_{i,j=1}^{3}\left\{ \bar{u}
_ {i}\,[V_{\sigma 2}^{*}\,(Y_{u}^{\text{diag}}\,K_{CKM})_{ij}]
\,P_L\,\tilde{\chi}
_{\sigma}^{+}\,(\Gamma_D)^{ja}\,\tilde{d}_{a}-\bar{u}_{i}\,[g\,U_{\sigma
1}\,(K_{CKM})_{ij}]\, P_R\,
\tilde{\chi}_{\sigma}^{+}\,(\Gamma_D)^{ja} \,\tilde{d}_a\right.   \nonumber \\
& &  \left. +\,\bar{u}_{i}\,[U_{\sigma 2}\,(K_{CKM}\,Y_{d}^{%
\text{diag}})_{ij}]
\,P_R\,\tilde{\chi}_{\sigma}^{+}\,(\Gamma_D)^{(j+3)a}\,\tilde{d}_a
\right\} +\text{%
H.c.}
\end{eqnarray}
where the index $\sigma$ refers to chargino mass eigenstates.
$Y_{u,d}^{\text{diag}}$ are the
diagonal up- and down-quark Yukawa couplings, and $V$, $U$ are the usual
chargino rotation matrices defined by $U^{*}M_{\tilde {\chi}
^{+}}V^{-1}=\mathrm{diag}%
(m_{\tilde {\chi} _{1}^{+}},m_{\tilde {\chi} _{2}^{+}})$. The flavor
changing effects
arise from both the off-diagonal elements in the CKM matrix $K_{CKM}$
and from the soft supersymmetry breaking terms in $\Gamma_D$.

Finally, the relevant Lagrangian terms for the quark-up squark
neutralino ($\tilde {\chi}^{0}_n$) interaction are
\begin{eqnarray}
\mathcal{L}_{u\tilde{u}\tilde{\chi}^{0}}&=&\sum_{n=1}^{4}\sum_{i=1}^{3}\left\{
\bar{u}
_{i}\,N_{n1}^{*}\,\frac{4}{3}\frac{g}{\sqrt{2}}\tan \theta _{W} \,P_L\,\tilde{%
\chi}_{n}^{0}\,(\Gamma_U)^{(i+3)a}\,\tilde{u}_a-\bar{u}_{i}\,N_{n4}^{*}\,Y_{u}^{\text{%
diag}}\,P_L\,\tilde{\chi}_{n}^{0}\,(\Gamma_U)^{ia}\,\tilde{u}_a
\right.   \nonumber \\
&-& \left.\bar{u}_{i}\,\frac{g}{\sqrt{2}}\left( N_{n2}+%
\frac{1}{3}N_{n1}\tan \theta _{W}\right)\,P_R
\,\tilde{\chi}_{n}^{0}\,(\Gamma_U)^{ia}\,\tilde{u}_a%
-\bar{u}_{i}\,N_{n4}\,Y_{u}^{\text{diag}}\,P_R\,\tilde{\chi}_{n}^{0}\,(\Gamma_U)^{(i+3)a}\,%
\tilde{u}_a\right\} \,,\nonumber \\
\end{eqnarray}
where $N$ is the $4\times 4$ rotation matrix which diagonalizes the
neutralino mass matrix $M_{\tilde \chi^0}$, $N^{*}M_{\tilde
\chi^0}N^{-1}=\mathrm{diag}(m_{%
\tilde {\chi}_{1}^{0}},\,m_{\tilde {\chi}_2^0}, \,m_{\tilde
{\chi}_3^0}, \,m_{\tilde {\chi}_4^0})$. As in
the gluino case, FCNC terms arise only from supersymmetric parameters
in $\Gamma_U$.

Most of the previous analyses of
    FCNC processes in the MSSM concentrated on the  mass insertion
approximation \cite{Hall:1985dx}. In this formalism,
the ($\delta$) terms represent mixing between chirality states of
different squarks, and it is possible to compute the contributions of
the first order flavor changing mass insertions perturbatively, if one
assumes smallness of the inter-generational mixing elements
($\delta$'s) when compared with the diagonal elements. However, when
the off-diagonal elements
in the squark mass matrix become large, the mass insertion
approximation is no longer valid  \cite{Harnik:2002vs,Besmer:2001cj}.
In the general mass eigenstate formalism,  the mass matrix in
Eq.~(\ref{eq:gammaudef}) (and the similar one in the down-sector) is
diagonalized and the flavor changing parameters enter into our
expressions through the matrix $\Gamma_{U,D}$. So, in the rare top
decays $t\to cgg$, the new flavor changing neutral currents show
themselves in both gluino-squark-quark and neutralino-squark-quark
couplings in the up-type squark loops and in the
chargino-squark-quark coupling in the down-type squark loops.
      Therefore here, as in our previous work \cite{Frank:2005vd}, we use the
general mass eigenstate formalism as described above.

\section{$t \to cgg$ versus $t \to cg$ in MSSM}\label{sec:decays}
We present here the comparative analysis of the rare two and three body
top quark decays, $t \to
cgg$ and $t \to cg$,  closely following
the discussion in our earlier paper \cite{EFTSM}. There, we have
    shown that, within the SM framework, the branching
ratio of $t \to cgg$ is about two orders of magnitude larger than
that of $t \to cg$ in SM, a phenomenon which can be dubbed "higher
order dominance", and which was
revealed e.g., in $b$ and $c$-physics in
the past. For the detailed discussion, see
\cite{EFTSM} and the relevant references therein. Even though the branching
ratio for $t \to cgg$ dominates the one for the two body decay $t \to
cg$, it is of the order of $10^{-9}$ and still too small to be
detected in collider experiments. Any experimental signal for such
decay would indicate physics beyond the SM. So, our aim in this section
is to extend the discussion in \cite{EFTSM} to a favorable beyond SM
framework in which we would expect larger contributions due to extra
sources of FCNC -- the unconstrained MSSM. Note that we include the SM
contributions as well in our calculations.

\begin{figure}[htb]
\vspace*{-3.8in}
        \centerline{ \epsfxsize 9in {\epsfbox{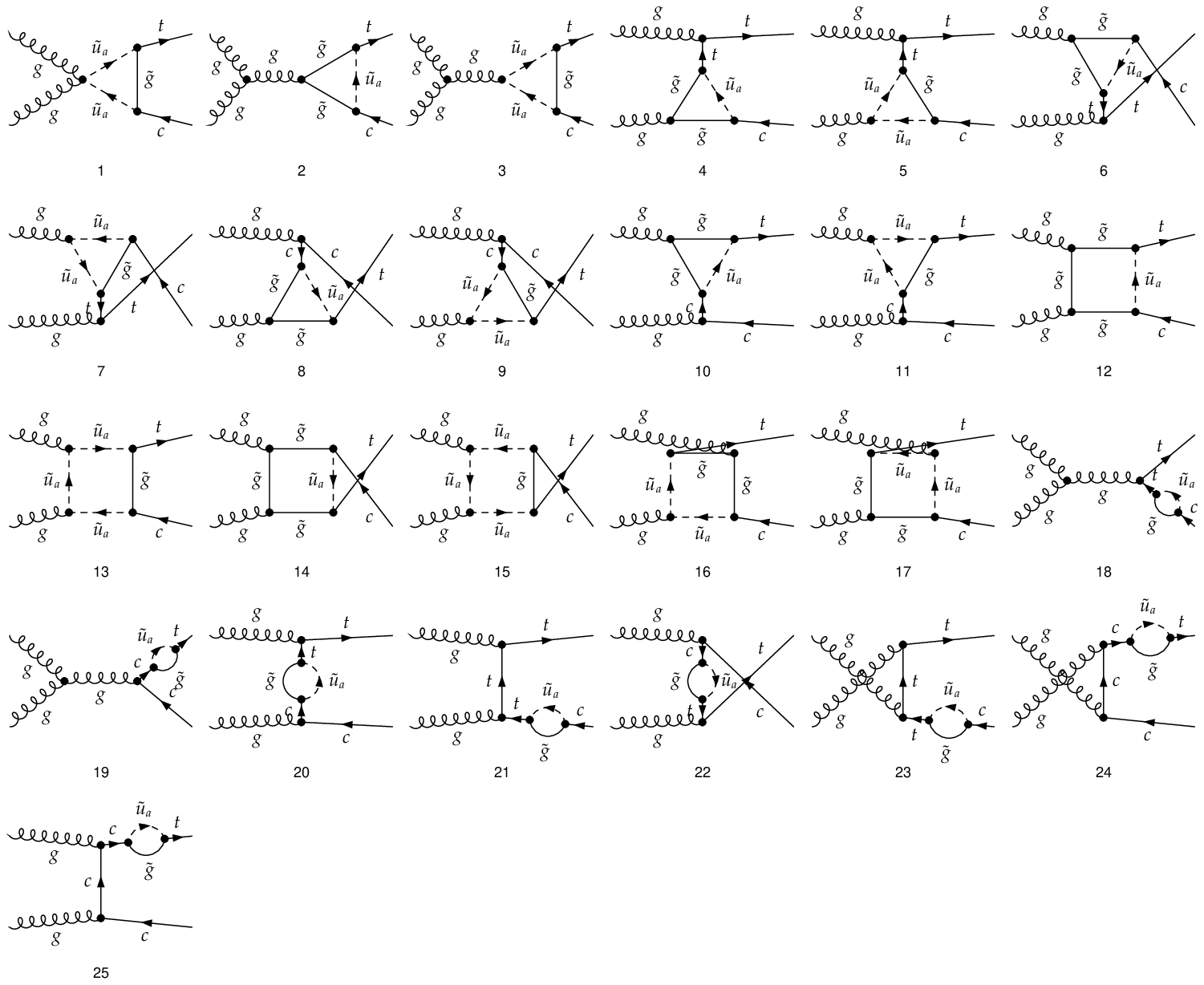}}}
\vspace*{-3.7in}
\caption
      {\texttt{The one-loop gluino contributions to $gg \to t\bar{c}$ in the
unconstrained MSSM in the
't Hooft-Feynman gauge.}}
\label{fig:gluino}
\end{figure}

\begin{figure}[htb]
\vspace*{-3.8in}
        \centerline{ \epsfxsize 9in {\epsfbox{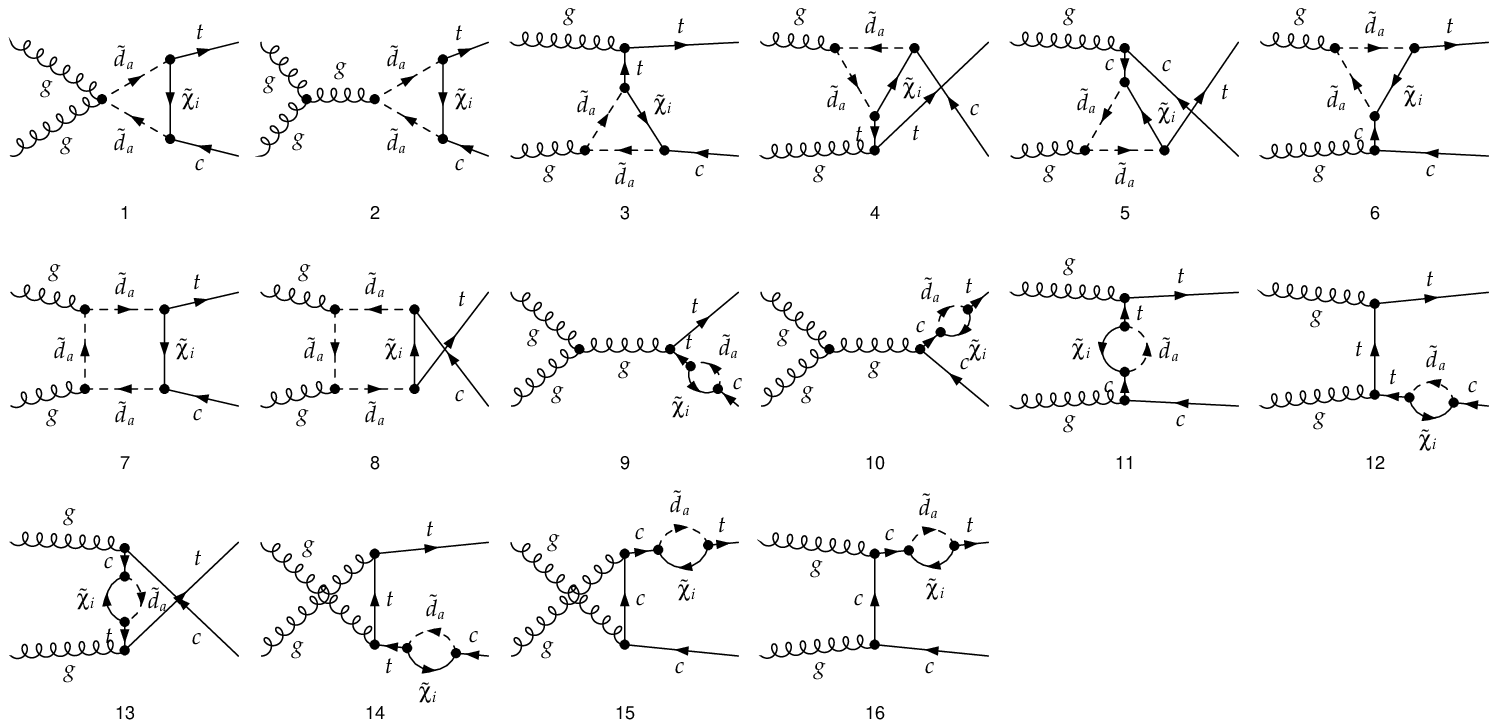}}}
\vspace*{-5.9in}
\caption
      {\texttt{The one-loop chargino contributions to $gg \to t\bar{c}$ in the
unconstrained MSSM in the
't Hooft-Feynman gauge.}}
\label{fig:chargino}
\end{figure}

\begin{figure}[htb]
\vspace*{-3.8in}
        \centerline{ \epsfxsize 9in {\epsfbox{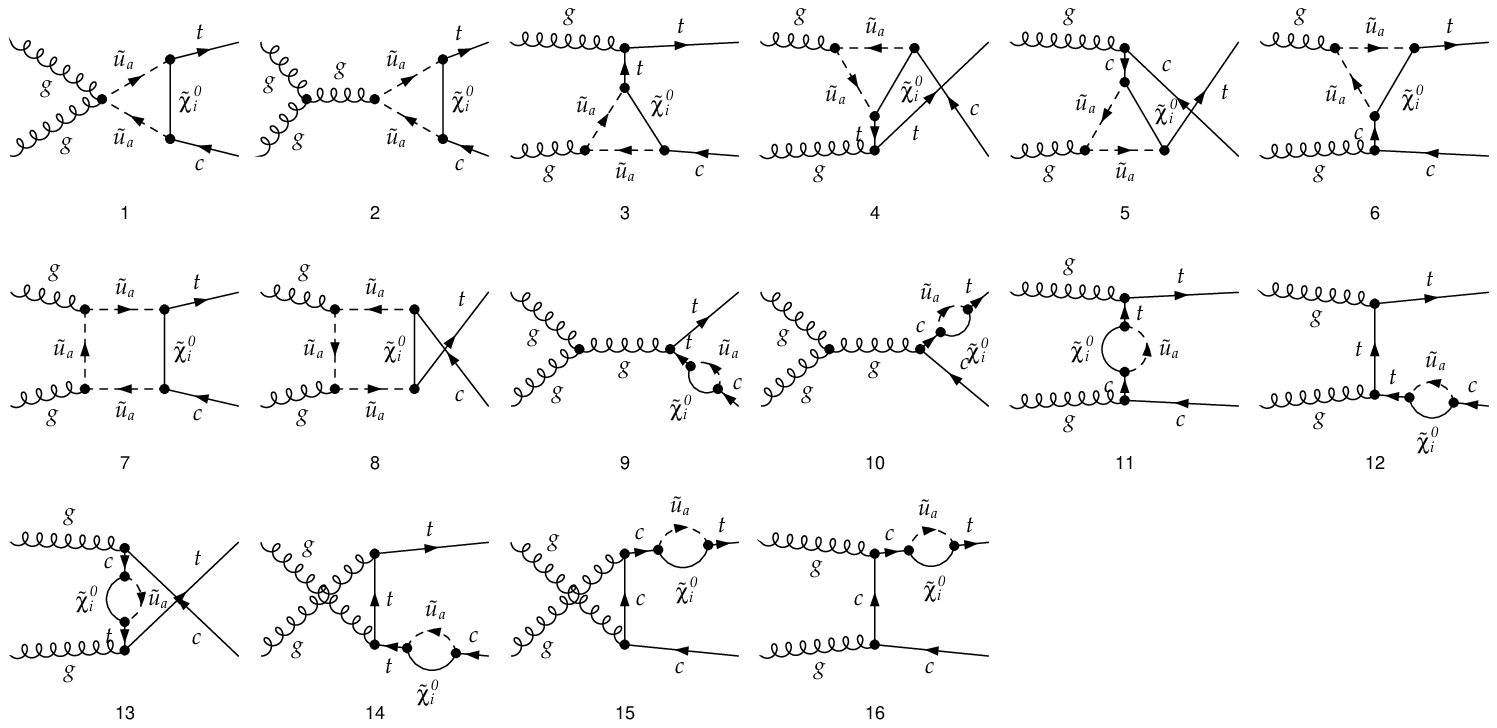}}}
\vspace*{-5.9in}
\caption
     {\texttt{The one-loop neutralino contributions to $gg \to
t\bar{c}$ in the
unconstrained MSSM in the
't Hooft-Feynman
gauge.}}
\label{fig:neutralino}
\end{figure}

\begin{figure}[htb]
\vspace*{-3.8in}

\centerline{ \epsfxsize 9in
{\epsfbox{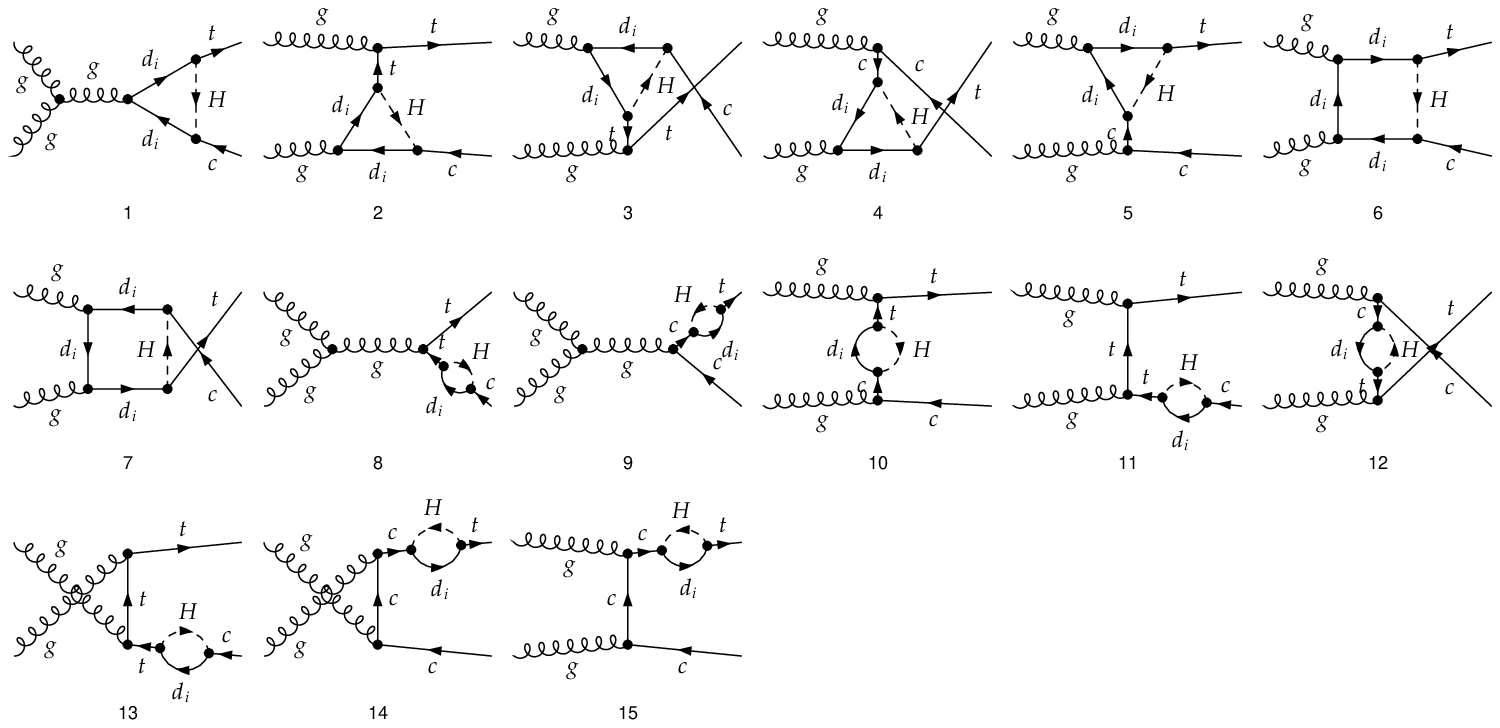}}}
\vspace*{-5.9in}
\caption
     {\texttt{The
one-loop charged Higgs contributions to $gg \to t\bar{c}$ in
the unconstrained MSSM in the
't Hooft-Feynman
gauge.}}
\label{fig:higgs}
\end{figure}

\begin{figure}[h]
\vspace*{-5in}

\centerline{ \epsfxsize 9in
{\epsfbox{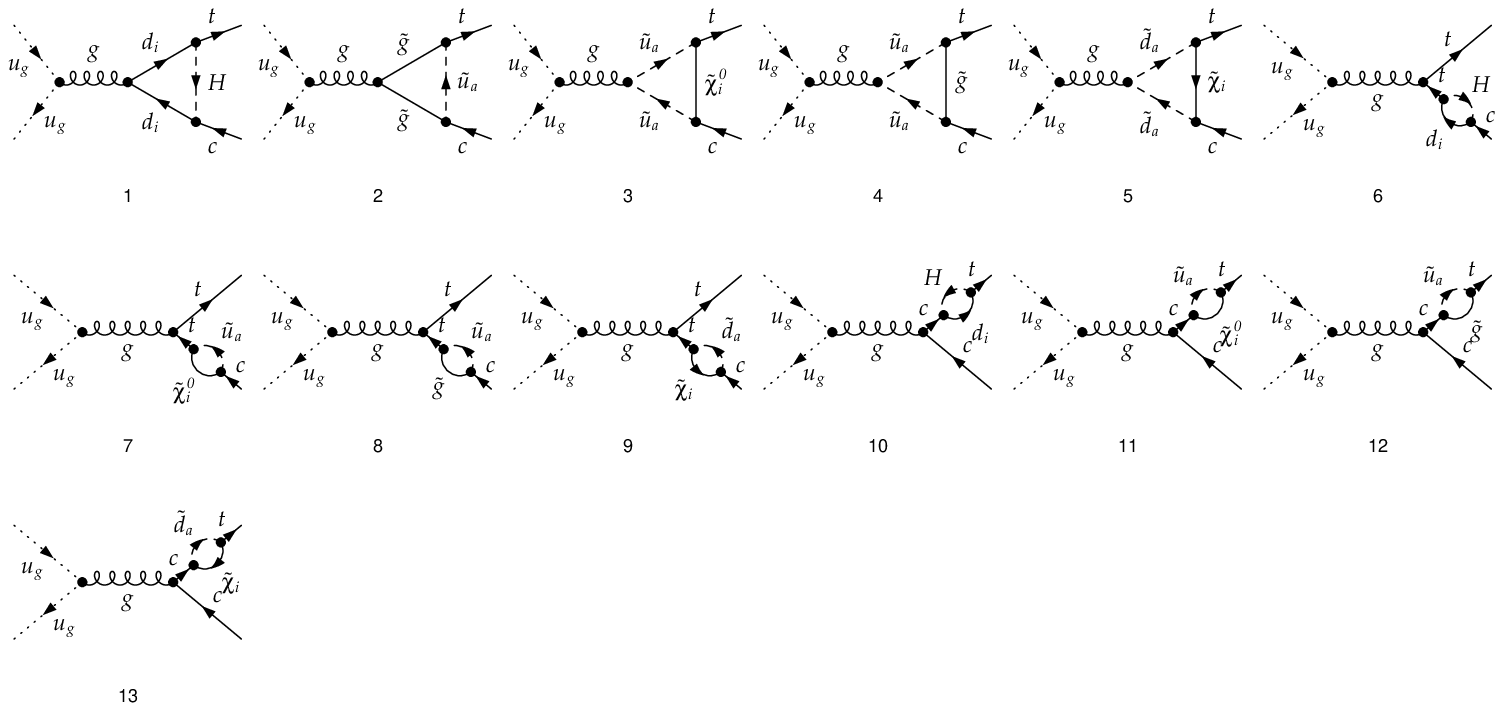}}}
\vspace*{-4.8in}
\caption
     {\texttt{The
one-loop QCD ghost contributions to $gg \to t\bar{c}$ in the
unconstrained MSSM in the
't Hooft-Feynman
gauge.}}
\label{fig:ghost}
\end{figure}

The one-loop Feynman diagrams contributing to $t \to cgg$ in the MSSM are
given in a set of diagrams
Figs.~\ref{fig:gluino}, \ref{fig:chargino}, \ref{fig:neutralino},
\ref{fig:higgs}, and \ref{fig:ghost} in the 't Hooft-Feynman gauge ($\xi = 1$)
representing gluino, chargino, neutralino, Higgs, and ghost
contributions, respectively.\footnote{Note that we display the
one-loop diagrams for the process $gg \to t\bar{c}$.
The diagrams for the decay can be easily obtained by crossing.}
We did not show the SM
diagrams here (since they appear in \cite{EFTSM}) but we took them into
account in the numerical
evaluation, for both the decays and the production mode.

As in \cite{EFTSM}, we choose to use the 't Hooft-Feynman gauge in which
the gluon polarization sum is
$\sum_{\lambda}\epsilon^*_{\mu}(k,\lambda)\epsilon_{\nu}(k,\lambda)=-g_{\mu\nu}$.
In order to obey unitarity, this simple choice results in the
existence of QCD ghost fields whose
contributions are shown in Fig.~\ref{fig:ghost}.  We closely follow
the method outlined in \cite{EFTSM} and references therein for
handling  the ghost diagrams.

Divergences inherent in the $t \to cgg$ calculation are ultraviolet,
infrared, and collinear types \cite{EFTSM}.  In numerical
evaluations, we used the softwares
\texttt {FeynArts}, \texttt{FormCalc}, and \texttt{LoopTools}
\cite{Hahn:2000jm} to obtain our results. In addition to these, \texttt{
HadCalc} \cite{hadcalc} is used for deriving the $pp$ process
corresponding to the $gg$ fusion discussed in the next
section.
Using
utilities offered by \texttt{ FormCalc}, we checked ultraviolet
finiteness of our results numerically, and introduced phase space cuts
to avoid infrared and collinear singularities.\footnote{These
cuts lead to some uncertainties in our results. A more precise
approach requires full consideration of the next-to-leading
order corrections to $t \to c g$, similar to the ones
in $b$ decays \cite{Greub:2000sy}.}

Having mentioned some qualitative features of the decay $t \to cgg$,
we do not present here most of the analytical intermediate results. We do this
since the calculations are lengthy and uninspiring.
Furthermore,
we use well known programs.\footnote{The complete analytical results
can be obtained by contacting one of us (I.T.)}
We have also checked our calculations with similar ones, whenever
published, as we discuss in the next section.

We express the matrix element squared $\left|{\cal M}\right|^2$
as a sum over the various contributions. These include the SM
contribution as given in our
previous work \cite{EFTSM}. From Figs.~\ref{fig:gluino},
\ref{fig:chargino}, \ref{fig:neutralino}, \ref{fig:higgs}, and
\ref{fig:ghost}, we obtain expressions
for the following non-SM terms: the gluino contribution, chargino,
neutralino, charged Higgs and finally the contribution of the ghosts.

The results were expressed in terms of Passarino-Veltman
functions \cite{Passarino:1978jh}. Numerical evaluations of these
functions have been carried out with \texttt{
LoopTools}, which does not require reduction of Passarino-Veltman
functions to the scalars
$A_0, B_0, C_0$ and $D_0$. The analytical expressions are obtained
with the use of \texttt{
FeynCalc} \cite{Mertig:1990wm}.

The partial width $d\Gamma$ for the decay
$t\to cgg$ is given as
\begin{eqnarray}
d\Gamma(t\to cgg) &=& \frac{1}{2m_t}\sum_{\rm{spins}}|{\cal M}|^2
d\Phi_3(k_1;k_2,k_3,
k_4)\nonumber\\
d\Phi_3(k_1;k_2,k_3,k_4) &=& \frac{d^3k_2}{(2\pi)^3 2k_2^0} \frac{d^3k_3}
{(2\pi)^3 2k_3^0} \frac{d^3k_4}{(2\pi)^3 2k_4^0} (2\pi)^4
\delta^{(4)}(k_1-k_2-k_3-k_4),
\end{eqnarray}
where $k_1(k_2)$ is the momentum of the top( charm) quark and
$k_3,k_4$ the momenta of the
gluon pair. The volume element can further be expressed as
\begin{eqnarray}
d\Phi_3(k_1;k_2,k_3,k_4) = \frac{1}{32\pi^3}\int_{(k_3^0)^{\rm min}}^{(k_3^0)
^{\rm max}}dk_3^0\int_{(k_2^0)^{\rm min}}^{(k_2^0)^{\rm max}}dk_2^0,
\end{eqnarray}
where the limits are
\begin{eqnarray}
(k_2^0)^{\rm min} &=& {\rm Max}\left[C m_t,\frac{\sigma-|{\bf
k}_3|}{2}\right],\nonumber\\
(k_2^0)^{\rm max} &=& \frac{\sigma+|{\bf
k}_3|}{2}(1-2C),\nonumber\\
(k_3^0)^{\rm min} &=& C m_t,\nonumber\\
(k_3^0)^{\rm max} &=& \frac{m_t}{2}(1-2C),
     \end{eqnarray}
      with $\sigma = m_t -k_3^0$. In addition, $C$ is the cutoff parameter,
    chosen nonzero to avoid infrared and collinear singularities
    \cite{EFTSM}.  For the numerical calculations in the rest of our study
we fix  $C= 0.1$, which is large enough to be able to reach the jet
energy  resolution sensitivity of the LHC detector. The results are sensitive to the choice of the $C$ parameter; we find that by decreasing $C$ to $0.01$, $BR(t \to cgg)$ can increase by a factor of 2-4. 

The total decay width of the top quark is taken to be $\Gamma_t=1.55$ GeV. The
parameters  used in our numerical evaluation are given in
Table~\ref{parameters}.
\begin{table}[htb]
	\caption{\texttt{The parameters used in the numerical
calculation.}}\label{parameters}
\begin{center}
        \begin{tabular}{c c c c c c}
        \hline\hline
$\alpha_s(m_t)$ &\hspace*{1cm} $\alpha(m_t)$ &\hspace*{1cm}
$\sin\theta_W(m_t)$ &\hspace*{1cm} $m_c(m_t)$ &\hspace*{1cm}
$m_b(m_t)$ &\hspace*{1cm} $m_t(m_t)$\\
\hline
      0.106829  &\hspace*{1cm}   0.007544    &\hspace*{1cm}  0.22
&\hspace*{1cm}  0.63 GeV  &\hspace*{1cm}  2.85 GeV &\hspace*{1cm}
174.3 GeV  \\
\hline \hline
\end{tabular}
\end{center}
\end{table}

The MSSM parameters $M_{\rm {SUSY}},\, M_2,\, m_{A^0},\, \mu, A$, and
$\tan\beta$ are chosen as
free for the constrained MSSM and
the SUSY-GUT mass relations are assumed.\footnote{The existence of a
GUT theory at Planck
scale leads to relations among
gaugino mass parameters of the form $$M_1=(5s_W^2/3c_W^2)M_2 =
(5\alpha/3c_W^2\alpha_s)m_{\tilde{g}}$$ where $\alpha$ and $\alpha_s$
are running coupling
constants.} (This is the first scenario we consider). Inclusion of
the flavor violating parameters $\delta$'s among second and third
generation squarks (the unconstrained MSSM) adds eight more free
parameters. Imposing SUSY-GUT
relations favors a heavy gluino, which decreases the gluino
contributions for both
processes under consideration, $t\to cg(g)$ and $gg\to t\bar{c}$, and
which enhances chargino
contributions, since the lightest chargino becomes much lighter than gluino.

As a second scenario we consider the constrained and unconstrained MSSM
without imposing SUSY-GUT relations. In this case,
we run the $U(1)$ gaugino mass parameter $M_1$ and the gluino mass
$M_{\tilde{g}}$
separately.\footnote{We still keep the relation between $M_1$
and $M_2$, rather than fixing them independently, since this
does not affect significantly the final results.} Thus the two
scenarios we
    concentrate on are MSSM with, and MSSM without, SUSY-GUT relations.

Given the still large number of parameters in either of these
scenarios, the parameter space
needs to be reduced by making further assumptions. So, for
simplicity, we assume
that the soft SUSY-breaking parameters in the squark sector are set to the
common value
$M_{\rm {SUSY}}$. In addition to this, the trilinear linear terms
$A_{u_i}$ and $A_{d_i}$ are
chosen to be real and equal to each other and $\mu$ is also taken to
be real and positive.

In the case of flavor violating MSSM, only the mixing between the second and
the third
generations is turned on, and the dimensionless parameters
$\delta$'s run over as much of  the interval
(0,1) as allowed.\footnote{Even though $\delta$'s are allowed to be
negative, we run them in the
positive region.} The allowed upper limits of $\delta$'s are
constrained by the requirement that
$m_{\tilde{u}_i,\tilde{d}_i}>0$ and  consistent with the experimental
lower bounds (depending on the chosen values of
$M_{\rm {SUSY}},
A, \tan\beta$, and $\mu$).  We
assume a lower bound of $96$ GeV for all up squark masses and $90$ GeV
for the down squark masses \cite{Eidelman:2004wy}. The Higgs
masses are calculated with
\texttt{FeynHiggs} \cite{feynhiggs}, with the requirement
that the lightest neutral
Higgs mass is larger than $114$ GeV.  Other experimental bounds
included are  \cite{Eidelman:2004wy}: $96$ GeV for the
lightest chargino,
$46$ GeV the lightest neutralino, and $195$ GeV for the gluino.
Throughout the paper, only
$m_{A^0}$ and $A$ are fixed globally in the decay and production
separately, $m_{A^0}=400$
GeV and $A=620$ GeV  in the decay process $t\to cgg$ (and $t\to cg$
as well) and
$m_{A^0}=500$ GeV and $A=400$ GeV, respectively, in the single top
production process $gg\to
t\bar{c}$.
\begin{figure}[htb]
\vspace{0.05in}
         \centerline{\hspace*{-0.1cm} \epsfxsize 3.1in
{\epsfbox{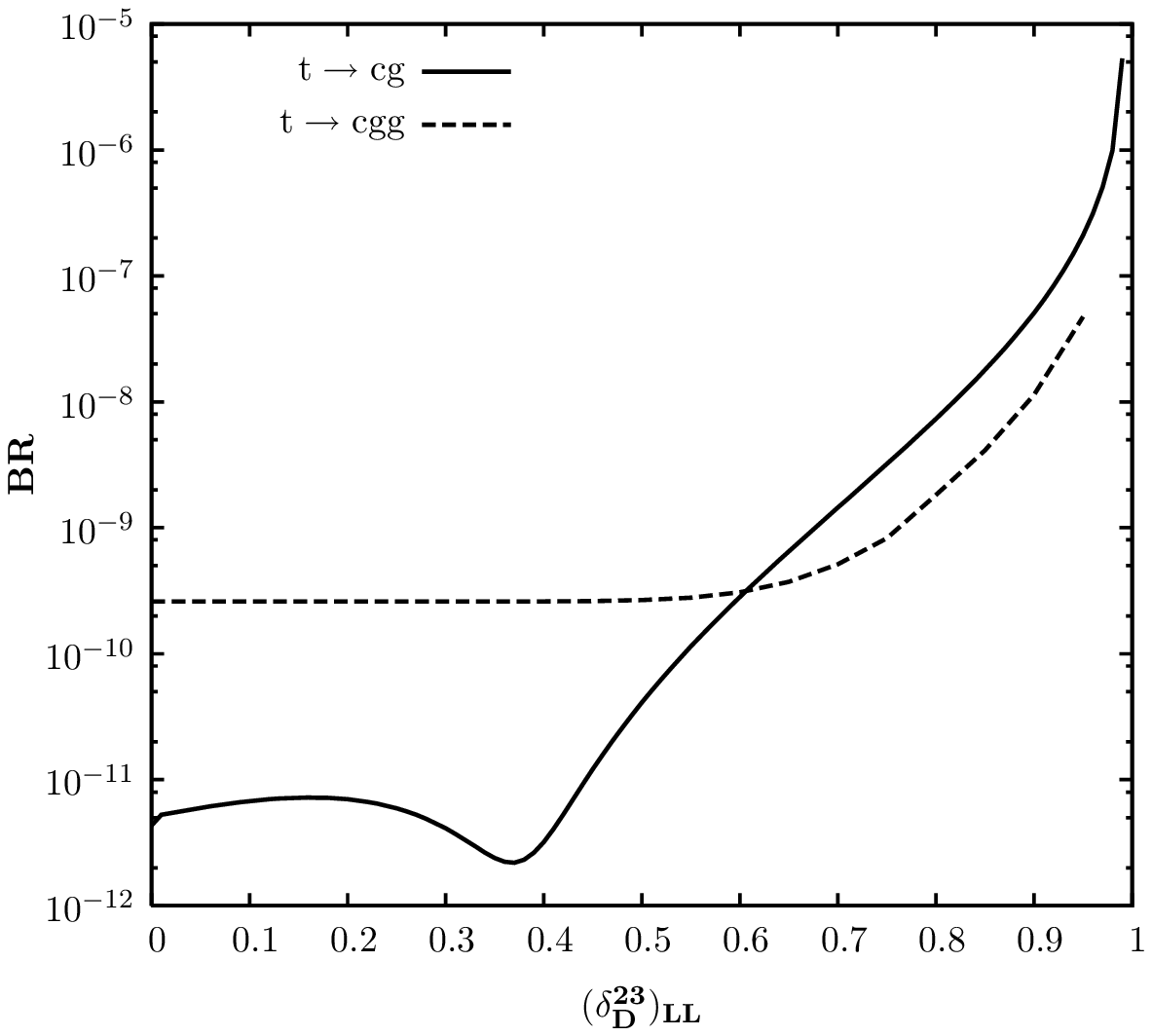}} \hspace{-0.01cm} \epsfxsize 3.1in
{\epsfbox{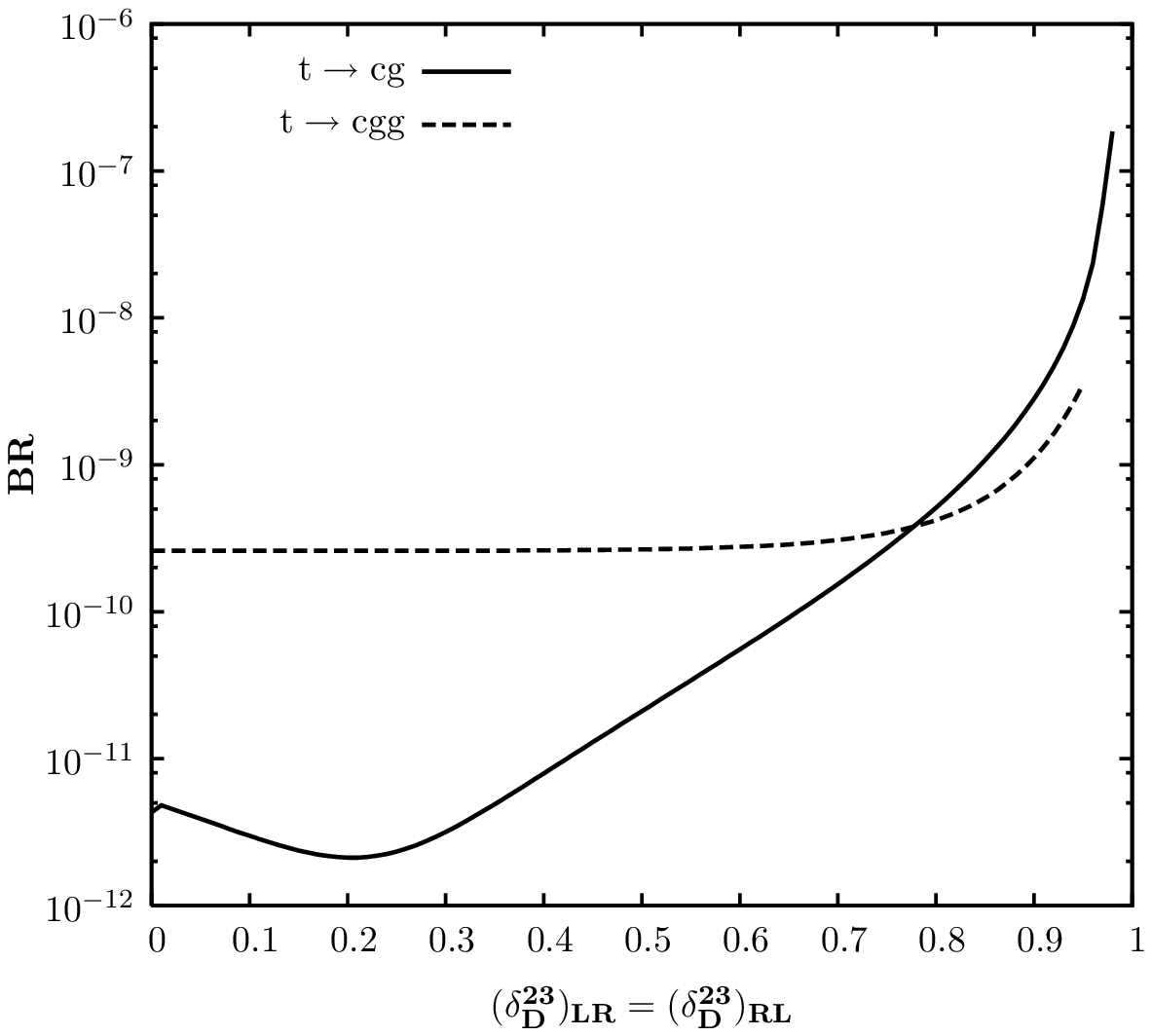}} }
\vskip -0.2in
        \caption{\texttt{Left panel: Branching ratios of
$t\to cgg$ and $t\to cg$ decays as
functions of $(\delta^{23}_D)_{LL}$ with the assumption that GUT
relations hold. Right panel:
Branching ratios as functions of
$(\delta^{23}_D)_{LR}=(\delta^{23}_D)_{RL}$
under the same conditions. The parameters are chosen as
$\tan\beta=10$, $M_{\rm {SUSY}}=300$
GeV, $M_2=\mu=200$ GeV.}}\label{fig:tcggD}
\end{figure}

\begin{figure}[htb]
\vspace{0.05in}
         \centerline{\hspace*{-1.3cm} \epsfxsize 3.1in
{\epsfbox{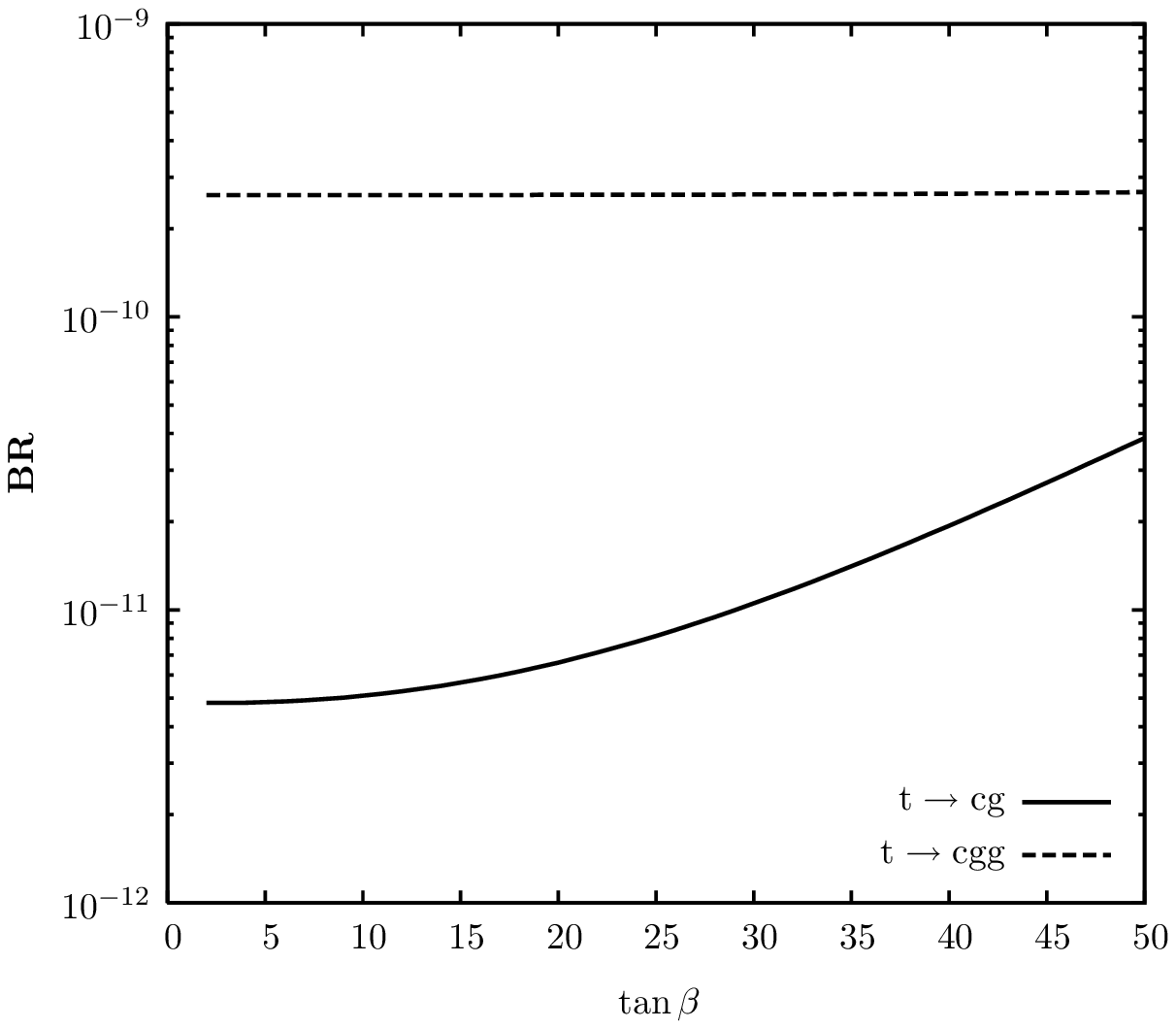}} }
\vskip -0.2in
         \caption{\texttt{The branching ratios of $t\to cgg$ and $t\to
cg$ decays as functions of
$\tan\beta$ with the assumption that GUT relations hold. It is
further assumed that all
flavor off-diagonal parameters $\delta$'s are zero in both the up and down
sectors (constrained MSSM). The other
parameters are chosen as $M_{\rm {SUSY}}=300$ GeV, $M_2=\mu=200$
GeV.}}\label{fig:tcggTB}
\end{figure}

\begin{figure}[htb]
\vspace{0.05in}
         \centerline{\hspace*{-0.1cm} \epsfxsize 3.1in
{\epsfbox{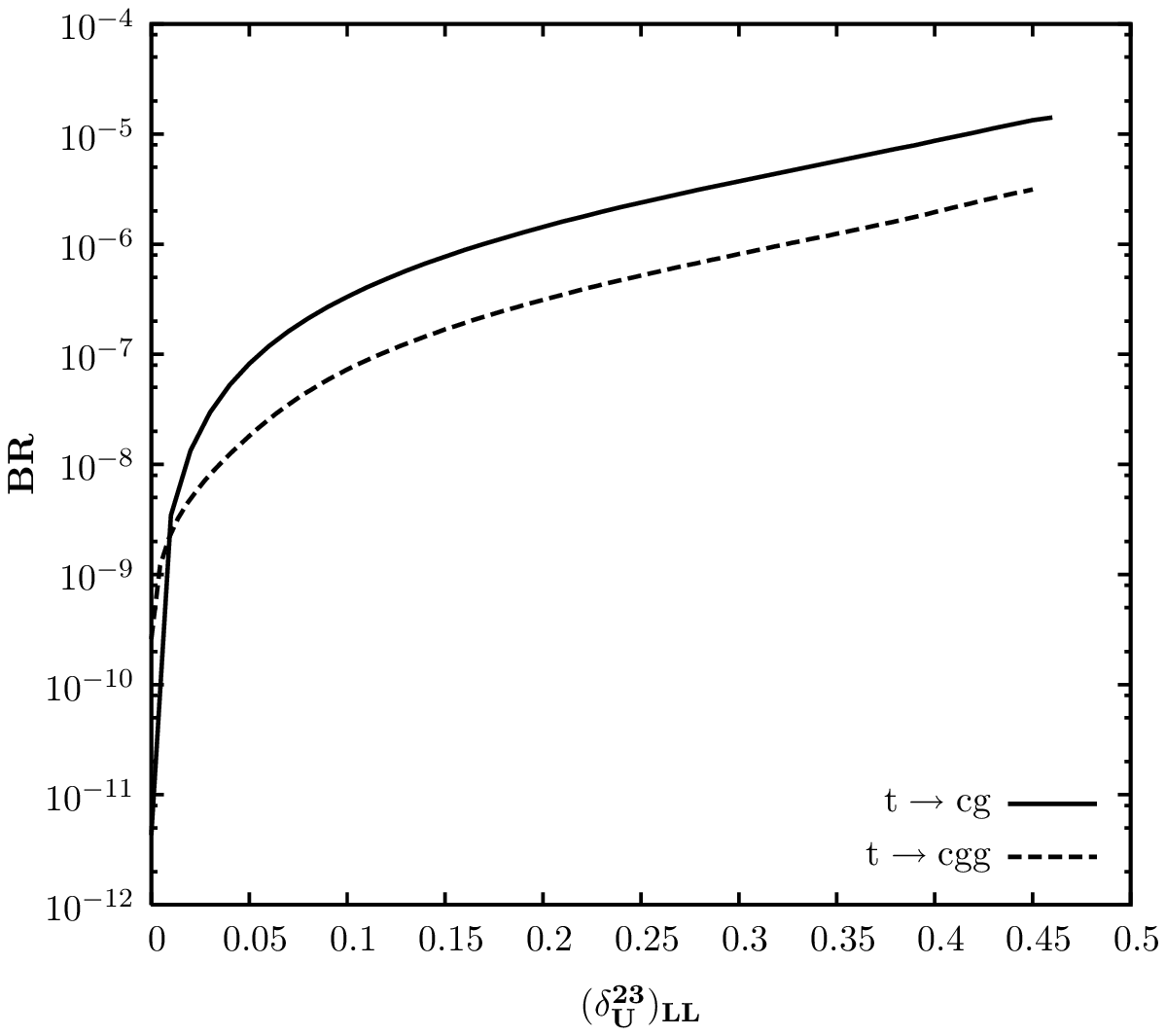}} \hspace{-0.01cm} \epsfxsize 3.1in
{\epsfbox{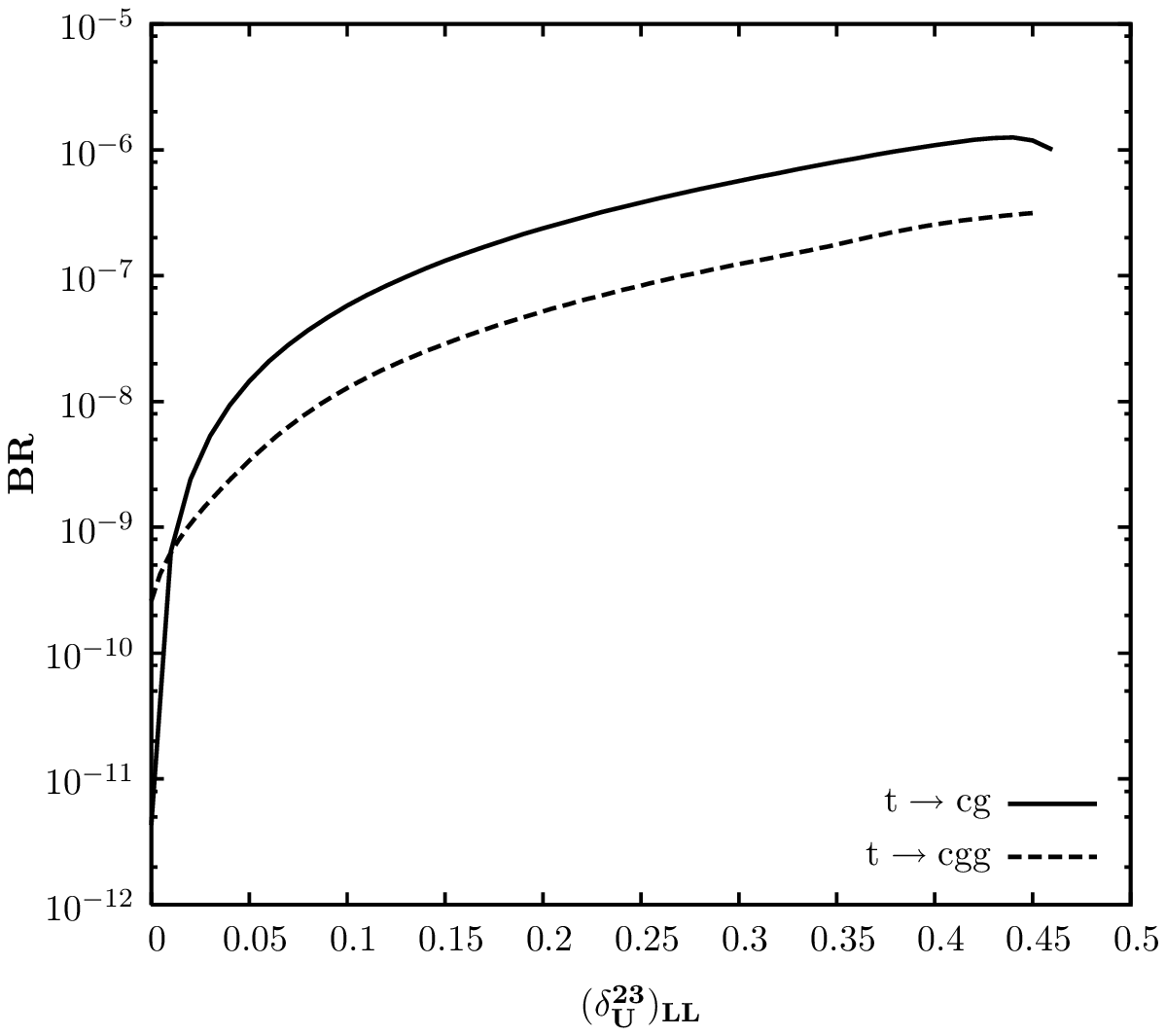}} }
\vskip -0.2in
         \caption{\texttt{The branching ratios of $t\to cgg$ and $t\to
cg$ decays as functions of
$(\delta^{23}_U)_{LL}$ without GUT relations for $m_{\tilde{g}}=200$
GeV, on the left panel,
and  $m_{\tilde{g}}=300$ GeV, on the right panel. The parameters are
chosen as $\tan\beta=10$,
$M_{\rm {SUSY}}=300$ GeV, $M_2=\mu=200$ GeV.}}\label{fig:tcggNoGUT}
\end{figure}

The rest of the section is devoted to the presentation of our results
for the three body decay $t\to cgg$
and the comparison with the two body channel $t\to cg$, both within the
MSSM framework.
Since the flavor violating parameters $\delta$'s play very
important role in both
decays (both are flavor-violating rare top decay
channels), we vary them by keeping
only a single flavor off-diagonal element non-zero unless otherwise stated. In
this section,
$\tan\beta = 10$ is chosen in all figures except for
Fig.~\ref{fig:tcggTB}, where
the dependence of the $\rm BR$'s on $\tan\beta$ are shown.
Furthermore, the common SUSY
scale $M_{\rm {SUSY}} = 300$ GeV; $M_2=200$ GeV, and $\mu=200$ GeV
are chosen and fixed
globally in this section. Since we are only interested in the relative
size of the $ BR(t\to
cgg)$ with respect to $ BR(t\to cg)$, we consider the scenario of
MSSM with GUT
gaugino mass relations for illustration purposes, and present the
case without GUT mass relations in one figure at
the end of the section, namely Fig.~\ref{fig:tcggNoGUT}.

Fig.~\ref{fig:tcggD} shows the branching ratios  of the
decays $t\to cgg$ and $t\to cg$ as functions of
$(\delta^{23}_D)_{LL}$ on the left panel, and as functions of
$(\delta^{23}_D)_{LR}=(\delta^{23}_D)_{RL}$ on the right panel. Since
the flavor off-diagonal $\delta$'s in the up sector are switched off,
these figures
show chargino-only contributions. As seen from the panels, $
BR(t\to cgg)$ is  almost
two orders of magnitude larger than
$ BR(t\to cg)$ in most of the parameter space, and especially for
small $\delta^{23}_D$, up to
$\delta^{23}_D\sim 0.4$. As
$\delta^{23}_D$'s become larger, $ BR(t\to cg)$ increases rapidly
and  becomes larger than $ BR(t\to cgg)$ for $(\delta^{23}_D)_{LL}\ge
0.6$
  for left-panel and  for $(\delta^{23}_D)_{LR}\ge 0.8$ for the right
panel.
  The maximum value reached is around $10^{-7}$
for non-zero
$(\delta^{23}_D)_{LL}$ and $10^{-8}$ for the special case
$(\delta^{23}_D)_{LR}=(\delta^{23}_D)_{RL}$. (Note that $t\to cg$ can get
even larger in this part of the phase space).
   These two figures
demonstrate explicitly
that $t\to cgg$ is larger than $t\to cg$ over most of the parameter
space. We have checked the
dependence of $ BR(t\to cgg)$  and $ BR(t\to cg)$ on
$(\delta^{23}_D)_{RR}$ and
observed that $ BR(t\to cgg)$ remains two orders of magnitude
larger than $ BR(t\to
cg)$ for the most part of the interval, while the sensitivity to
$(\delta^{23}_D)_{RR}$ variations is not as pronounced as in the
(depicted) $LL$ and
$LR,~RL$ cases. In this case,
$ BR(t\to cg)$ can reach a few times $10^{{-9}}$.
\begin{figure}[htb]
\vspace{0.05in}
         \centerline{\hspace*{-0.1cm} \epsfxsize 3.1in
{\epsfbox{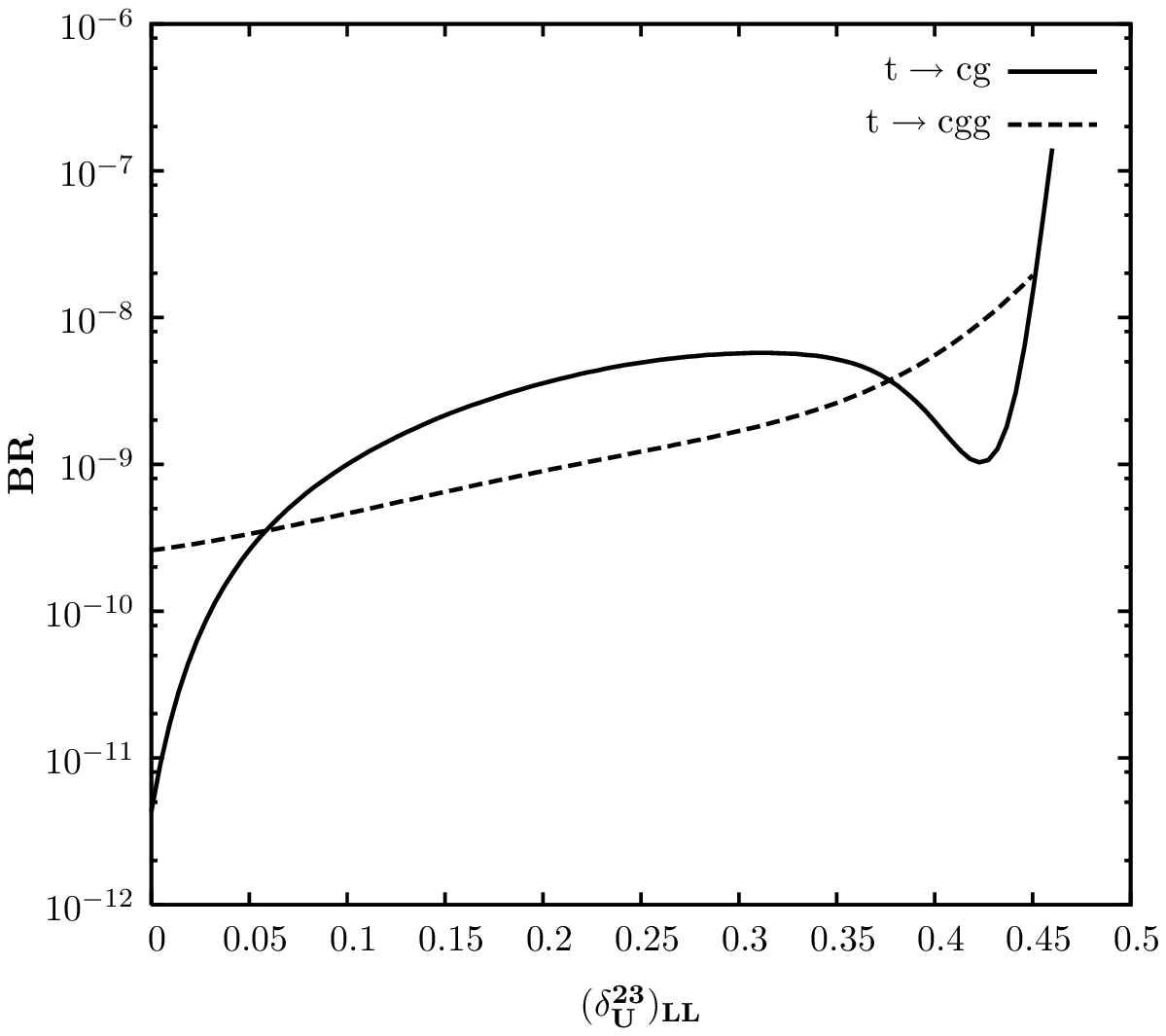}} \hspace{-0.01cm} \epsfxsize 3.1in
{\epsfbox{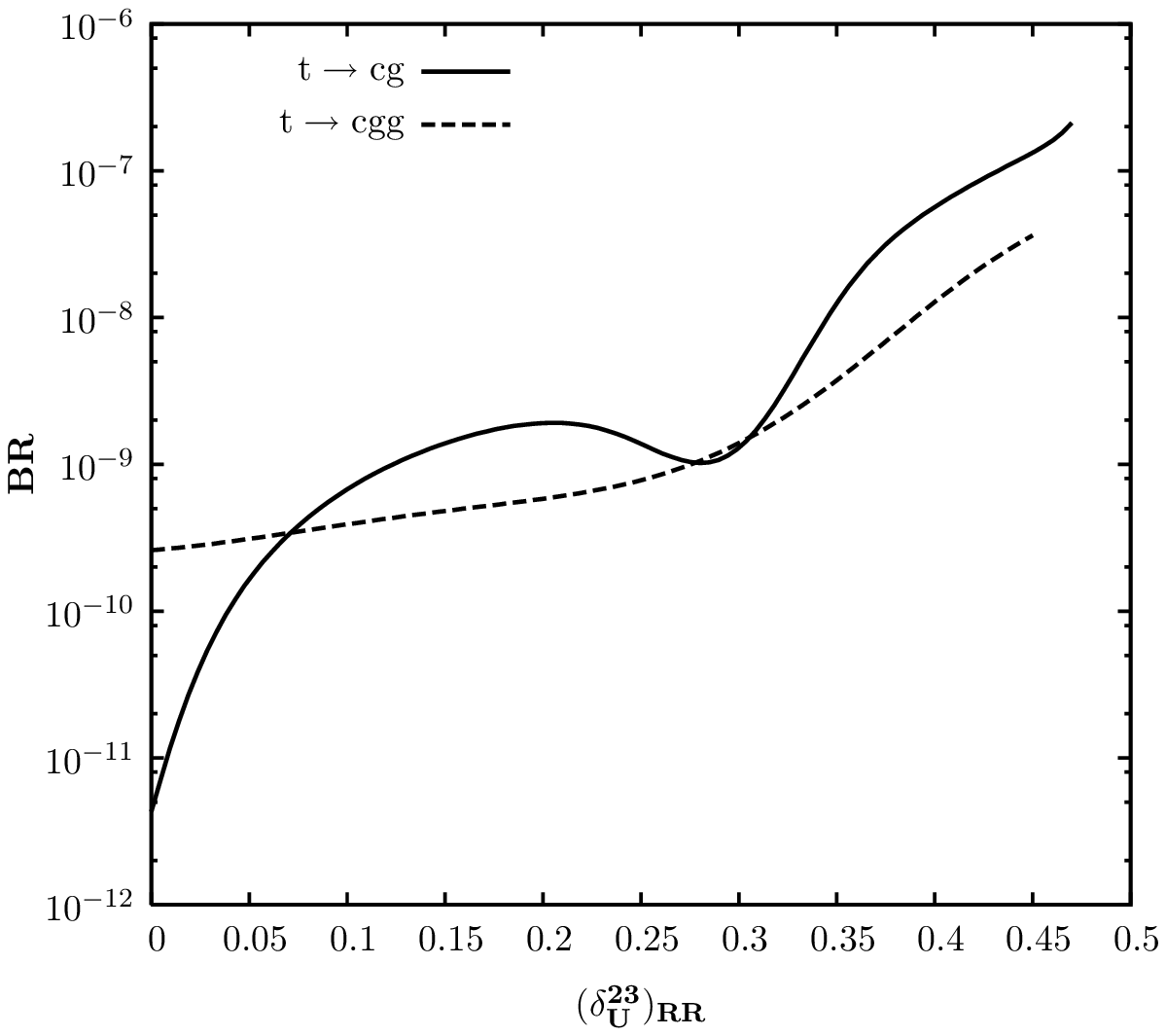}} }
\vskip -0.2in
          \caption{\texttt{Left panel: Branching ratios of
$t\to cgg$ and $t\to cg$ decays as
functions of $(\delta^{23}_U)_{LL}$ with the assumption that GUT
relations hold. Right panel:
Branching ratios as functions of $(\delta^{23}_U)_{RR}$
under the same
conditions. The parameters are chosen as $\tan\beta=10$, $M_{\rm
{SUSY}}=300$ GeV,
$M_2=\mu=200$ GeV.}}\label{fig:tcggU}
\end{figure}

In Fig.~\ref{fig:tcggU}, the $(\delta^{23}_U)_{LL}$ and $(\delta^{23}_U)_{RR}$
dependence of the branching ratios of $t\to cgg$ and $t\to cg$ decays
are
shown on the left and right panels, respectively. Since the GUT
relations are assumed to hold,
the gluino mass is rather heavy, about $600$ GeV, when $M_2$ is chosen as
$200$ GeV. The two
orders of magnitude difference between the $\rm BR$'s for the flavor conserving
MSSM disappear once we
introduce a small flavor violation ($\sim 0.1$) between the second and
third generations in
the up squark sector, which holds for either  $LL$ or $RR$ case.  The
branching ratio of
$t\to cg$  exceeds that of $t\to cgg$ for $\delta^{23}_U\ge 0.1$. The
maximum attainable branching ratio for
$t\to cg$ is around $10^{-7}$, and for $t\to cgg$, $10^{-8}-10^{-7}$  
which represents two orders of
magnitude enhancement for
$t\to cgg$, and more than $4$ orders of magnitude enhancement for
$t\to cg$, with respect to the
constrained case. The case of
$(\delta^{23}_U)_{LR}=(\delta^{23}_U)_{RL}$ is very similar to the case
with non-zero $(\delta^{23}_U)_{LL}$ (left panel) or
$(\delta^{23}_U)_{RR}$ (right panel).

Fig.~\ref{fig:tcggTB} shows the $\tan\beta$ dependence of the decays with
zero flavor off-diagonal parameters $\delta=0$ for $M_{\rm {SUSY}}=300$ GeV,
$M_2=\mu=200$ GeV. For the decay $t\to cgg$, the SUSY contribution
comes from the chargino sector in the
   MSSM (there are no gluino or neutralino contributions.)
     Overall the  SM contribution dominates over the MSSM one and the 
$\tan\beta$
dependence is insignificant, as
expected, since the constrained MSSM gives smaller contributions
than the SM to FCNC decays
at one-loop level. There is a mild dependence on $\tan\beta$ for $t\to cg$
decay in the very large $\tan\beta$ region ($\ge 25$). In addition to
that, we analyzed the case
with non-zero
$\delta$'s as well and, for example, for $(\delta_U^{23})_{LL}=0.4$,
we obtain $ BR(t\to
cgg)$ almost two orders of magnitude larger than $ BR(t\to cg)$ in the entire
$\tan\beta$ interval
considered.

The last figure of the section, Fig.~\ref{fig:tcggNoGUT}, presents
the dependence of the branching ratios on the SUSY flavor-violating
parameters in the MSSM without
SUSY-GUT relations. For illustration, we present the
$(\delta_U^{23})_{LL}$ dependence of the
$BR$'s for the gluino mass $m_{\tilde{g}}=200$ GeV on the left panel,
and for $m_{\tilde{g}}=300$
GeV on the right panel. The other parameters are chosen the same as
before, $M_{\rm {SUSY}}=300$
GeV, $M_2=\mu=200$. As seen from the figure, the relative difference
between the decays
not only disappears immediately after switching $(\delta_U^{23})_{LL}$ on
(more precisely, for $(\delta_U^{23})_{LL}\ge 0.01$) but
also $t\to cg$ exceeds $t\to cgg$ with a constant factor of 5. This is a
gluino dominated case which favors the two-body decay $t\to cg$ over the
three body decay.  The decay $t\to cg$ can get as large as
$10^{-5}$ for $m_{\tilde{g}}=200$ GeV and $10^{-6}$ for
$m_{\tilde{g}}=300$ GeV.

  From the analysis in this section, it is fair to say that the
branching ratio for the three
body $t\to cgg$ decay dominates largely over the one for the two body
$t\to cg$ mode for the flavor
conserving MSSM scenario with
SUSY-GUT relations, and remains larger even if non-zero
flavor off-diagonal parameters in the
  down squark sector are turned on.  Such dominance is
  valid only for
relatively small flavor violating parameter in the up
squark sector ($(\delta_U^{23})_{LL} < 0.1$).
  Our
results here show that the $t\to
cgg$ channel gives a larger contribution (and may be easier to
access) than $t\to cg$
channel over most of the parameter space if the flavor
violation originated from the down squark sector.

The predictions of the constrained MSSM
(without intergenerational squark mixings)  are similar to the
SM ones. Thus the existence of such SUSY FCNC mixings, directly related
to the  SUSY breaking mechanism, is crucial for the enhancement of the
branching ratios.

Another motivation for
considering $t\to cgg$ is the issue of single top quark production,
which is one of today's
challenging task at colliders. If $t\to cgg$ is a promising channel
with respect to
$t\to cg$ \footnote{The observability of $t\to cgg$ at LHC will be
briefly discussed at the end of the Section \ref{sec:production}. }, the next
question  would be what are the consequences of this for the single
top quark searches at colliders. For this purpose,
$gg\to t\bar{c}+\bar{t}c$ needs to be considered.
   Gluons will become very important and abundant at the
LHC, which reaches very
high energies. Therefore, the rest of the paper is devoted to the
consideration of the $pp\to
t\bar{c}+\bar{t}c+X$ cross section at LHC, within the flavor-violating MSSM, by
assuming only the gluon fusion contribution at partonic level.

\section{$pp \to t\bar{c}+\bar{t}c+X$ at LHC}\label{sec:production}

Having discussed the decay mode $t \to cgg$ and shown that it is a
more promising signal than $t \to cg$ in the previous section,
we consider here the top-charm associated production  via
gluon fusion $gg\to t\bar{c}+\bar{t}c$ at the partonic level. Since, at
the LHC, TeV or even higher-scale energies are going to be probed,
gluons inside the proton will become very important. This process, as
well as other channels involving light quarks,  has been considered
by Liu et. al \cite{Liu:2004bb} in the unconstrained MSSM driven by
SUSY-QCD contributions only. Their results show clearly
that $t\bar{c}$ production through gluon fusion is the dominant
channel over the ones involving light quarks
$q\bar{q}^{\prime},\,\,q,q^{\prime}=u,c,d,s$. For example, $87\%$ of
the total hadronic cross section $\sigma(pp\to t\bar{c}+\bar{t}c+X)$
comes from the partonic channel $gg\to t\bar{c}+\bar{t}c$ for
$(\delta_U^{23})_{LL,RR}=0.7$ \cite{Liu:2004bb}. We
agree with their results presented in \cite{Liu:2004bb} once we make
the required
modifications to the input parameters.

Here, we present the complete calculation of the hadronic cross
section $\sigma(pp\to t\bar{c}+\bar{t}c+X)$ at LHC by including all
one-loop contributions. In addition to the gluino, the chargino,
neutralino, and charged Higgs loops as well as the SM part is
included. The full set of Feynman diagrams contributing to the
process at one-loop level through gluino, chargino, neutralino, and
Higgs loops is given respectively in
Figs.~\ref{fig:gluino},~\ref{fig:chargino},~\ref{fig:neutralino}, and
\ref{fig:higgs} in the 't Hooft-Feynman gauge. As mentioned in the
previous section, we did not display here the SM diagrams available in our
previous paper \cite{EFTSM} for the $t \to cgg$ decay case. Note that, as
mentioned before, working in
the 't Hooft-Feynman gauge for this process requires the inclusion of
QCD ghost diagrams,  represented  in Fig.~\ref{fig:ghost}.

The partonic level
differential cross section for $gg\to t\bar{c}$  can be
expressed as
\begin{eqnarray}
d{\hat{\sigma}} &=&
\frac{1}{32\pi^2\hat{s}^{3/2}}|p|_{\rm {out}}|{\cal M}|^2
d\Omega_3,\nonumber\\
|p|_{\rm {out}}^2 &=&
\frac{(\hat{s}+m_t^2)^2}{4\hat{s}}-m_t^2
\end{eqnarray}
where $\Omega_3$ is
the angular volume of the third particle and $\sqrt{\hat{s}}$ is the
partonic center of mass energy.\footnote{For simplicity we assumed
$m_c$ zero in our analytical, but not in numerical, estimates.} The
matrix squared $|{\cal M}|^2$ can be
calculated by using the expressions,  for $t\to cgg$
by simply using the {\it crossing symmetry} (see
for example \cite{peskin}). Then, the hadronic cross section is
obtained by convoluting the partonic cross section with the parton
distribution functions (PDF's), $f_{g/p}$. So, the total hadronic cross
section reads
\begin{eqnarray}
\sigma =\int_{\xi_0}^1\,d\xi
\frac{d{\cal L}}{d\xi}\hat{\sigma}(\xi s,
\alpha_s\left(\mu_R\right))
\end{eqnarray}
where $\hat{\sigma}(\xi
{s}, \alpha_s\left(\mu_R\right))$ is the total partonic cross section
at the center of mass energy $\sqrt{\hat{s}}=\sqrt{\xi s}$
($\sqrt{s}$ is the hadronic center of mass energy) depending on the
renormalization scale $\mu_R$. Here $\xi_0$ defines the production
threshold of the process. The parton luminosity is defined as

\begin{eqnarray}
\label{limunosity}
\frac{d{\cal
L}}{d\xi}=\int_{\xi}^1\frac{dx}{x}f_{g/p}(x,\mu_F)f_{g/p}(\xi/x,\mu_F)
\end{eqnarray}
where
$\mu_F$ is the factorization scale, which is assumed to be equal to
the renormalization scale $\mu_R$ in our numerical analysis. If one
needs to sum over all
possible partonic subprocesses contributing to the particular final
state, there will be a sum over PDF's in  Eq.~(\ref{limunosity}).

We
assume that the top quark in the final state will be reconstructed
from events and thus it is a physical observable. Of course, to
identify the hadronic final state requires making a series of cuts on
the transverse momentum $p_T$ of the top and charm quarks, the
rapidity $\eta$,  and the jet separation $\Delta R_{34}$. The
following set is used for the
cuts
\begin{eqnarray}
p_{T_c},p_{T_t}\ge 15\, {\rm
GeV}\;\;\;\;\eta_c,\eta_t\le 2.5\;\;\;\;\Delta R_{34} \ge 0.4.
\end{eqnarray}
Their
effect is  translated into cuts on the limits of $\xi$ in
the calculation of the partonic cross section. For the transformation
of the initial partons to initial hadrons,
the program \texttt{HadCalc} \cite{hadcalc} was used, incorporating the
Les Houche Accord Parton Density Function library (LHAPDF) version
4.2 \cite{Giele:2002hx} with the recent data set CTEQ6AB
\cite{Pumplin:2005rh}.

For the numerical calculations, we have
chosen as input parameter the values  $m_{A^0}=500$ GeV, $A=400$ GeV.
In addition the hadronic
center of mass energy $\sqrt{s}=14$ GeV is taken for the LHC. The factorization
and renormalization scales  are chosen as the production threshold of
the process
($\mu_F=\mu_R=174.93$ GeV).

We
discuss the dependence of the total hadronic cross section of $gg\to
t\bar{c}+\bar{t}c$ process, $\sigma(pp\to t\bar{c}+\bar{t}c+X)$, on
various MSSM parameters for certain $\delta$ values in scenarios with
and without GUT relations. Note that for simplicity we assume a
common $\delta$ parameter in the up and down sector $(\delta_{U,
D}^{23})_{LL}=(\delta_{U, D}^{23})_{RR}=(\delta_{U,
D}^{23})_{LR}=(\delta_{U, D}^{23})_{RL}$ and only one (U {\it or} D)
non-zero at a time.  At the end of this section we discuss the relative
magnitude of the contributions coming from the gluino, chargino, and
the rest.
\begin{figure}[htb]
\vspace{0.05in}
\centerline{\hspace*{-0.9cm} \epsfxsize 3.5in
{\epsfbox{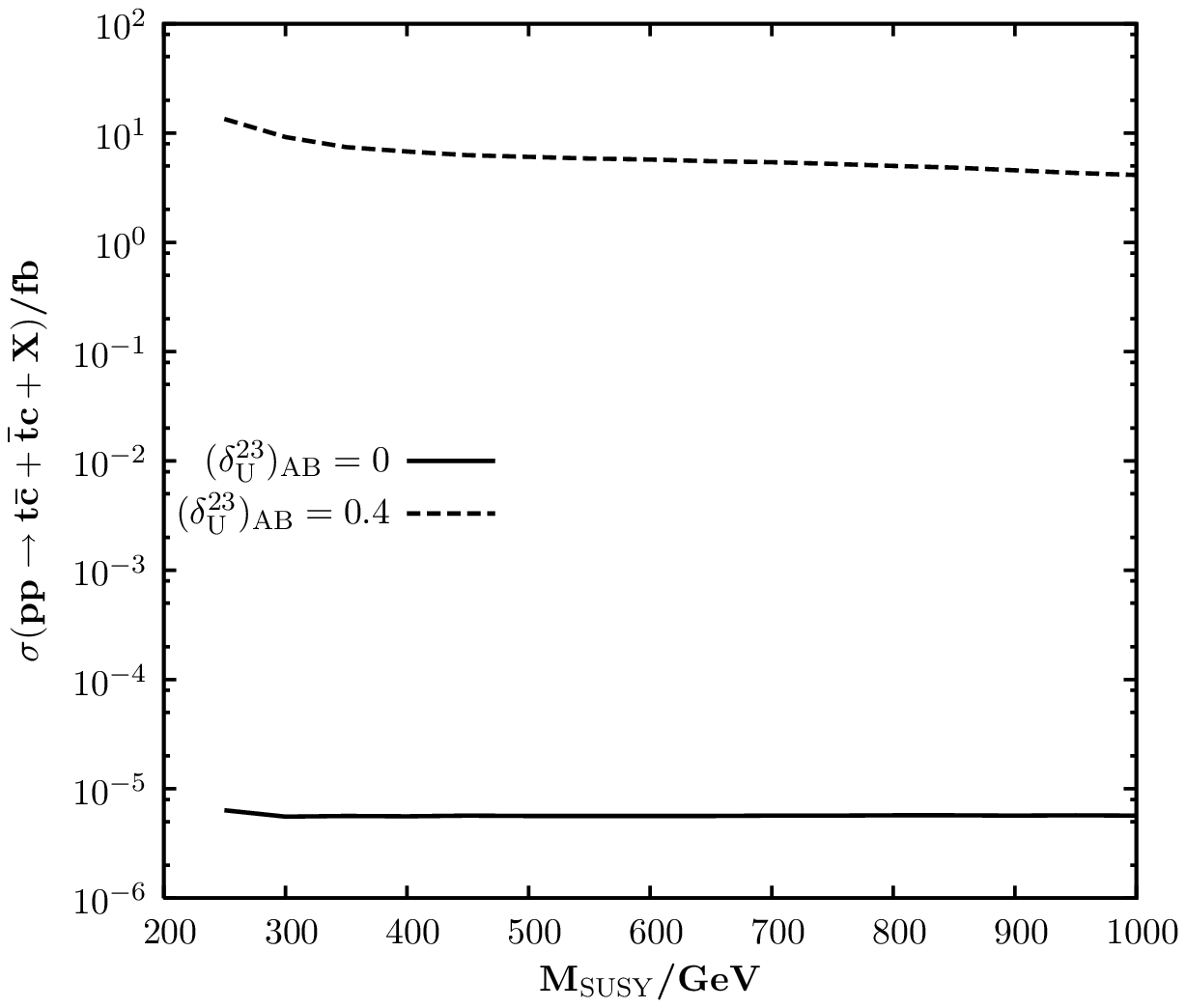}} \hspace{-1.2cm} \epsfxsize 3.5in
{\epsfbox{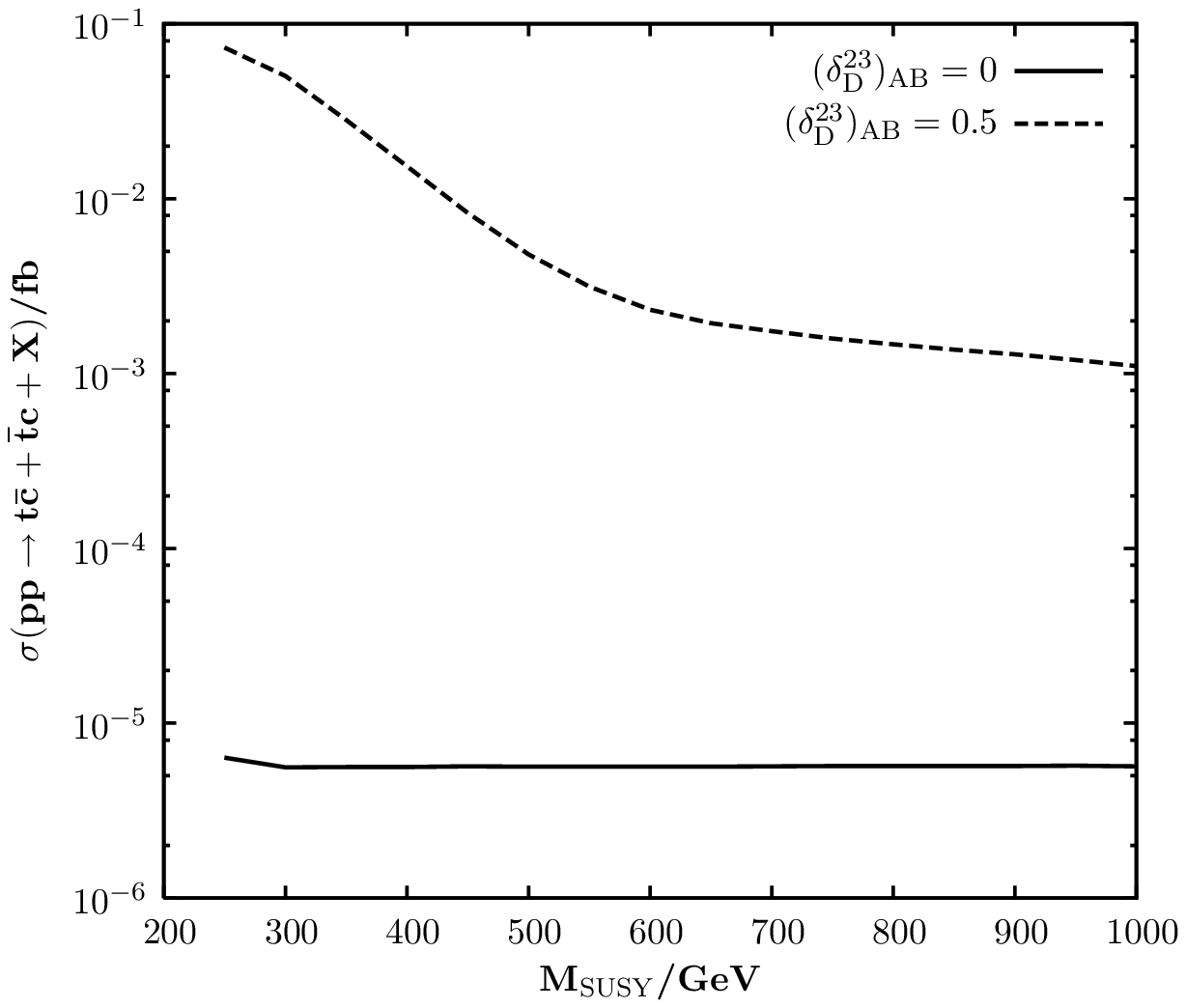}} }
\vskip -0.2in
\caption{\texttt{The total hadronic cross section $\sigma(pp\to
t\bar{c}+\bar{t}c+X)$ via gluon fusion as a function of $M_{\rm
{SUSY}}$ for $\tan\beta=5$, $\mu=250$ GeV, $M_2=200$ GeV, and
$m_{\tilde{g}}=300$ GeV. On the left panel, $\delta_U^{23}=0.4$ is
chosen and compared with the constrained MSSM case $\delta_U^{23}=0$.
The same is
shown on the right panel for $\delta_D^{23}=0.5$.}}
\label{fig:ggtcMsusy}
\end{figure}
\begin{figure}[htb]
\vspace{0.05in}
\centerline{\hspace*{-0.9cm} \epsfxsize 3.5in
{\epsfbox{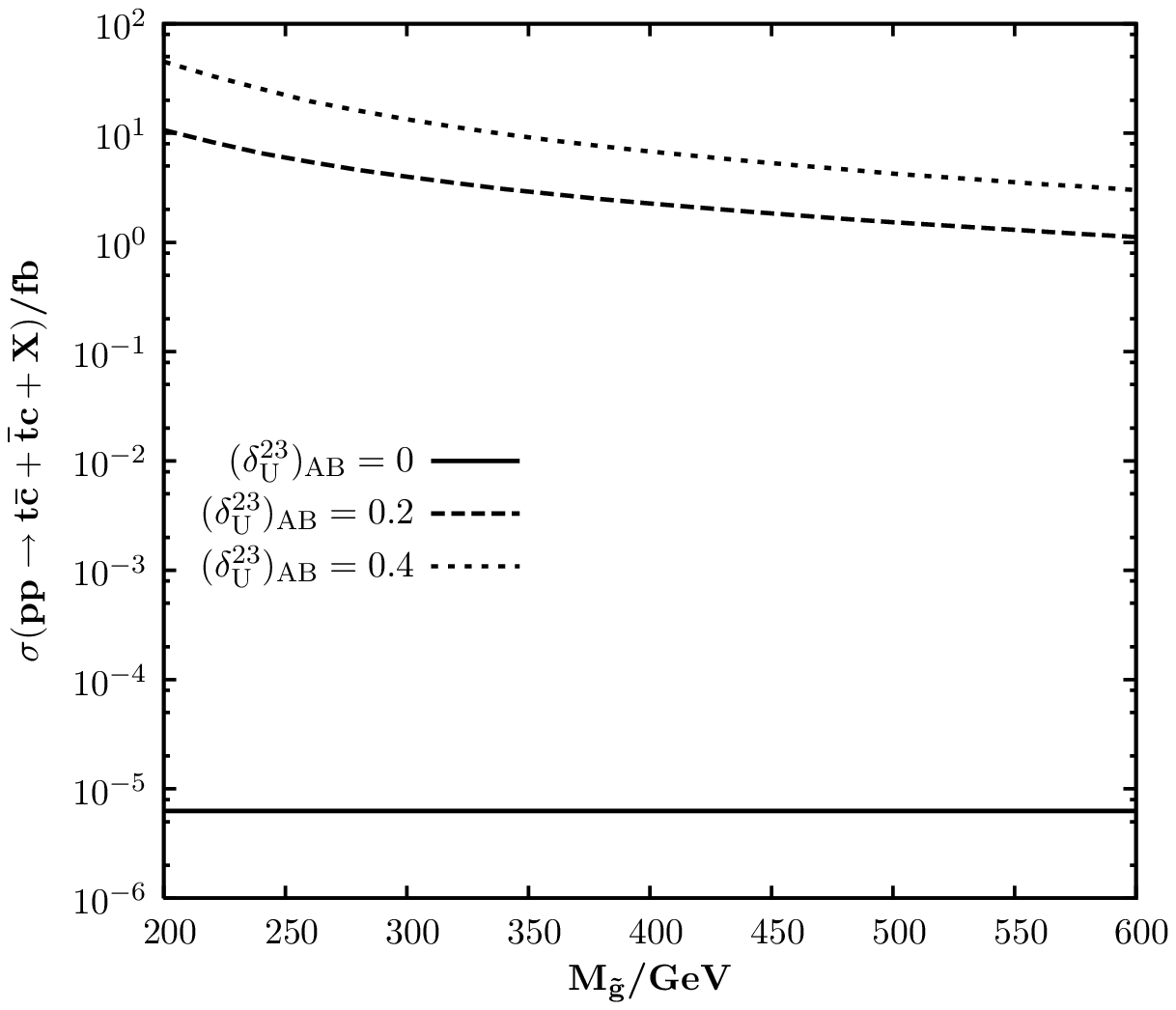}} \hspace{-1.2cm} \epsfxsize 3.5in
{\epsfbox{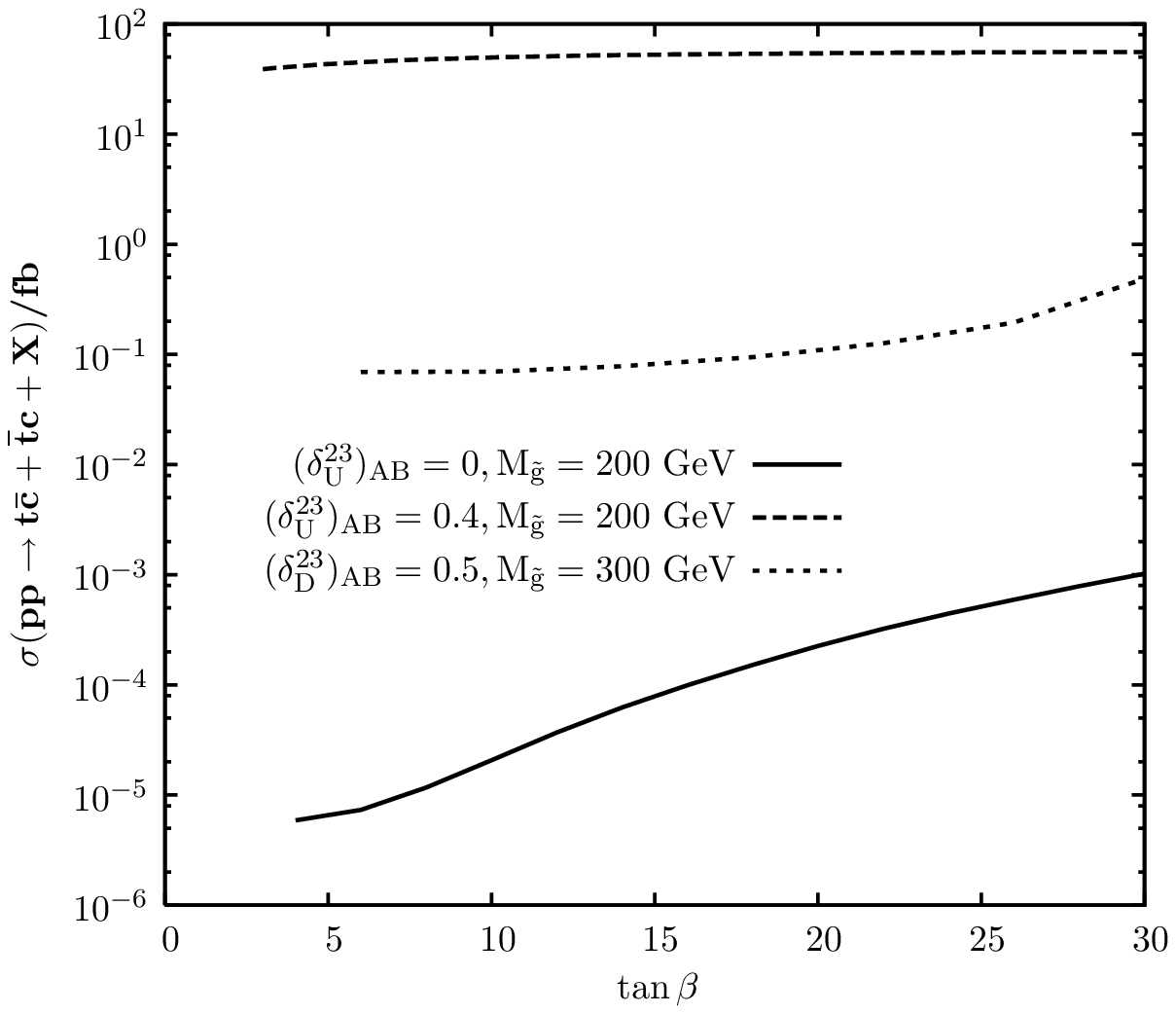}} }
\vskip -0.2in
\caption{\texttt{On the left panel, the total hadronic cross section
$\sigma(pp\to t\bar{c}+\bar{t}c+X)$ via gluon fusion as a function of
$m_{\tilde{g}}$ for $\tan\beta=5$, $M_{\rm {SUSY}}=\mu=250$ GeV, and
$M_2=200$ GeV at various $\rm (\delta_U^{23})_{AB}, ~A,B =L,R$. On
the right panel, $\sigma(pp\to t\bar{c}+\bar{t}c+X)$ as a function of
$\tan\beta$ at various $\delta_{U,D}^{23}$ and $m_{\tilde{g}}$
values.}}\label{fig:ggtcMGlTB}
\end{figure}
\begin{figure}[htb]
\vspace{-0.1in}
\centerline{\hspace*{-0.9cm} \epsfxsize 3.5in
{\epsfbox{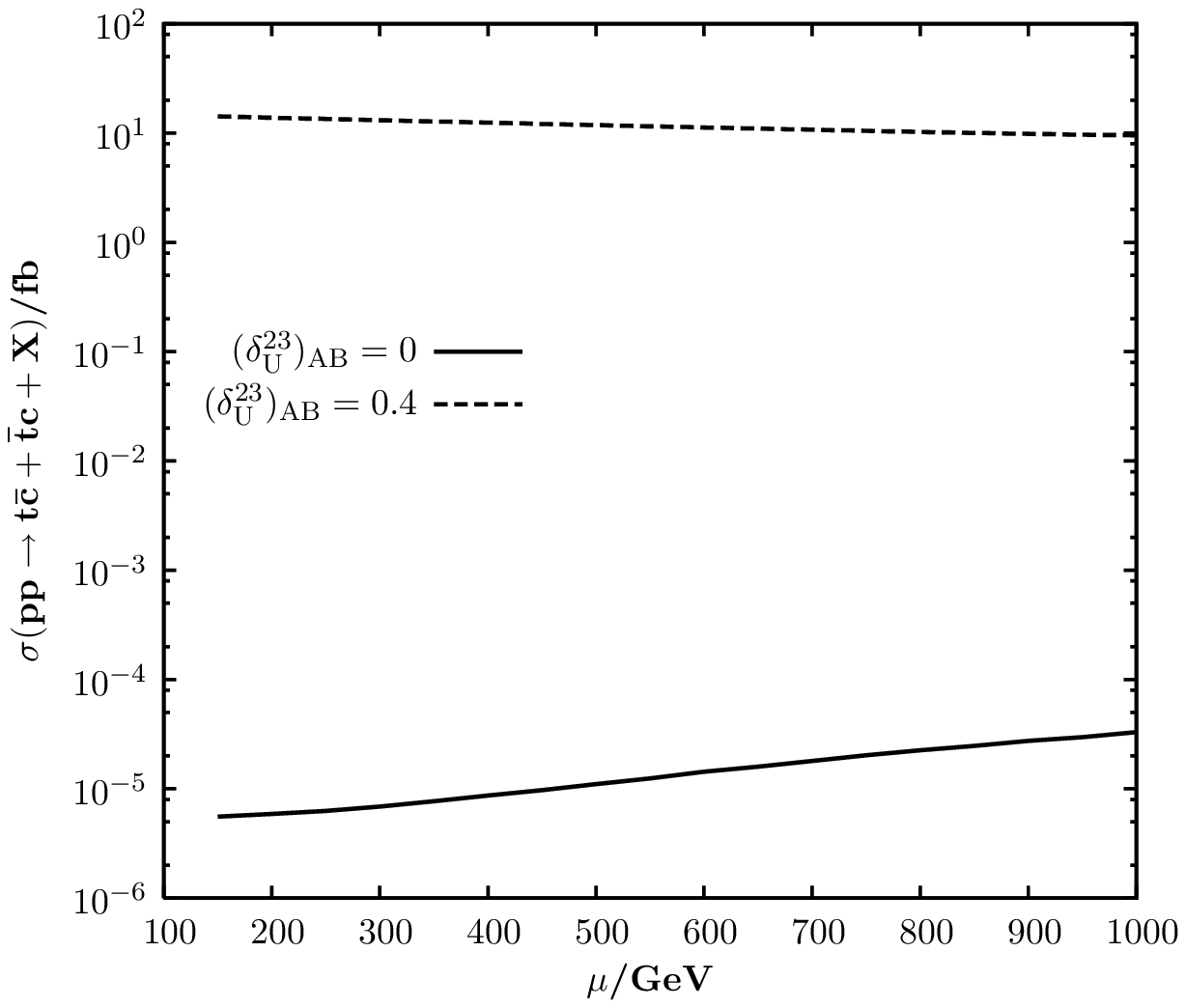}} \hspace{-1.2cm} \epsfxsize 3.5in
{\epsfbox{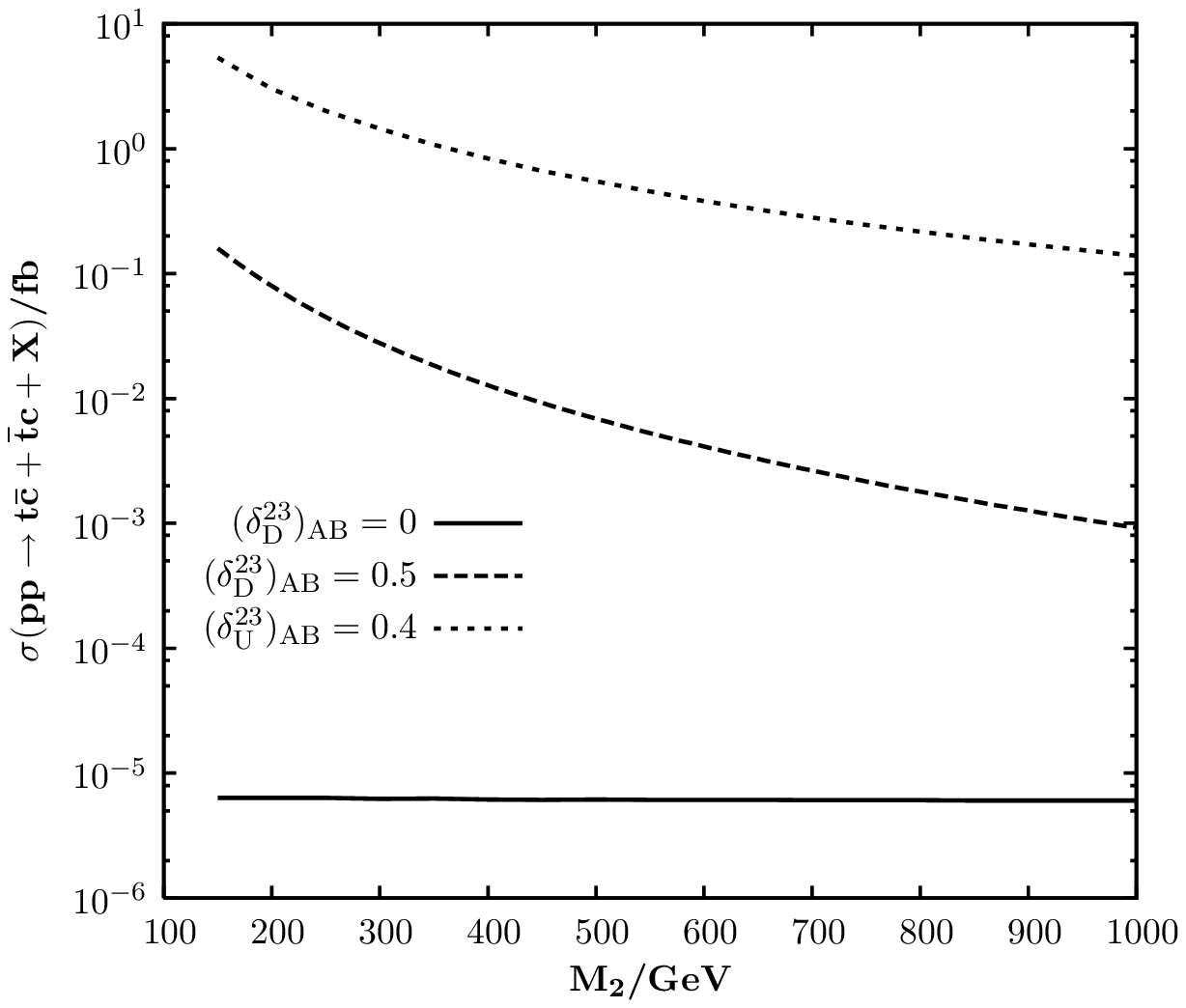}} }
\vskip -0.2in
\caption{\texttt{On the left panel, the total hadronic cross section
$\sigma(pp\to t\bar{c}+\bar{t}c+X)$ via gluon fusion as a function of
$\mu$ for $\tan\beta=5$, $M_{\rm {SUSY}}=250$ GeV, $M_2=200$ GeV,
and $m_{\tilde{g}}=300$ GeV at various $\rm (\delta_U^{23})_{AB}, A,B
=L,R$. On the right panel, $\sigma(pp\to t\bar{c}+\bar{t}c+X)$ as a
function of $M_2$ at various $(\delta_{U,D}^{23})_{AB}$ with GUT
mass relations.}}\label{fig:ggMUEM2}
\end{figure}

Fig.~\ref{fig:ggtcMsusy} shows
the $M_{\rm {SUSY}}$ dependence of the total hadronic cross section
$\sigma(pp\to t\bar{c}+\bar{t}c+X)$  for  $\tan\beta=5$,
$\mu=250$ GeV, $M_2=200$ GeV, and $m_{\tilde{g}}=300$ GeV. On the
left panel, there are two curves, for $(\delta_U^{23})_{AB}=0$, and
$0.4$. The cross section depends very weakly on $M_{\rm {SUSY}}$ and,
for the constrained MSSM case, the SM contribution is the dominant
one. There is an enhancement of more than 6
orders of magnitude in the unconstrained MSSM over the
constrained one, and the cross section can be as large  as $15$ fb
for $(\delta_U^{23})_{AB}=0.4$. In the down sector, shown on the
right panel, the sensitivity of $\sigma$ to $M_{\rm {SUSY}}$ is quite
strong, and there is an enhancement of about two orders of magnitude
in the interval $250-1000$
GeV. There are still $2-4$ orders of magnitude difference between the
constrained MSSM  versus the unconstrained MSSM scenarios at
$(\delta_D^{23})_{AB}=0.5$. The maximum cross section is around $0.1$
fb at around $M_{\rm {SUSY}}\sim 250$ GeV.

In
Fig.~\ref{fig:ggtcMGlTB}, on the left panel, the total cross section
$\sigma(pp\to t\bar{c}+\bar{t}c+X)$ is shown as a function of the
gluino mass for various $(\delta_U^{23})_{AB}$ values. Again the
constrained MSSM case is dominated by the SM contribution, while for
the unconstrained MSSM,  $\sigma\sim 45$ fb for $m_{\tilde{g}}=200$
GeV and $(\delta_U^{23})_{AB}=0.4$, which is more than 7 orders of
magnitude larger than for the case with $(\delta_U^{23})_{AB}=0$. On the
right panel, the $\tan\beta$ dependence of the total cross section is
shown for $\mu=250$ GeV, $M_2=200$ GeV, and $m_{\tilde{g}}=200,300$,
$(\delta_U^{23})_{AB}=0, 0.4$, and $(\delta_D^{23})_{AB}=0.5$. For
very large $\tan\beta$ values, the cross section reaches 0.001 fb in
the constrained MSSM, while in the unconstrained case, for
$(\delta_U^{23})_{AB}=0.4$ and $m_{\tilde{g}}=200$ GeV, a cross
section of $60$ fb is obtained. For $(\delta_U^{23})_{AB}=0.5$ and
$m_{\tilde{g}}=300$ GeV, the cross section under these conditions
reaches a few fb.

   Fig.~\ref{fig:ggMUEM2} illustrates the $\mu$ (on the left
panel) and $M_2$ (on the right panel) dependences of the total
hadronic cross section $\sigma$ for representative values of
$(\delta_{U,D}^{23})_{AB}$. The parameters are $\tan\beta=5$, $M_{\rm
{SUSY}}=250$ GeV, $M_2=200$ GeV, and $m_{\tilde{g}}=300$ GeV for the
left panel, and
$\tan\beta=5$, $M_{\rm {SUSY}}=\mu=250$ GeV, and
$m_{\tilde{g}}=300$ GeV for the right panel.
For non-zero
$(\delta_{U}^{23})_{AB}$,
the cross-section $\sigma$ is not sensitive to $\mu$ and
remains around $15$ fb,
but it decreases significantly with $M_2$ in
the interval $M_2 \in [150-1000]$ GeV, if there is a non-zero $\delta$
in either the up or down sector. The cross section ranges between 10 fb
to 0.1 fb for $M_2=150$ GeV and $1000$ GeV, respectively, for
$(\delta_{U}^{23})_{AB}=0.4$, and between $0.1$ fb to $0.001$ fb for
$(\delta_{D}^{23})_{AB}=0.5$.
\begin{table}[htb]
	\caption{\texttt{Relative contributions to the total cross
section $\sigma(pp\to t\bar{c}+\bar{t}c+X)$, in fb, with and without GUT
mass relations.
$(\delta_{U}^{23})_{AB}=(\delta_{D}^{23})_{AB}=0,\,0.2,\,0.4,\,\,A,B=L,R$
is considered. The rest of the parameters are $A=400$ GeV,
$\tan\beta=10$, $M_{\rm
{SUSY}}=\mu=250$ GeV and $M_2=200$ GeV. For the case without GUT,
$m_{\tilde{g}}=300$ GeV is used and $(\delta_{U,D}^{23})_{AB}$ are
given in brackets.}}
	\label{compare}
\begin{center}
    \begin{tabular*}{1\textwidth}{@{\extracolsep{\fill}} l l l l}
\hline\hline
$(\delta_{U}^{23})_{AB}=(\delta_{D}^{23})_{AB}\;\;$  & 0
& 0.2 (No GUT) &  0.4 (No GUT)\\
\hline\hline
    Gluino loop &  0  &
1.09 (4.05)    & 3.07 (14.13)  \\
Chargino loop &    $1.77 10^{-5}$
&      0.0052                        &         0.034 \\
The rest
&     $6.10 10^{-6}$      &      0.0025                       &
0.017  \\
\hline
\hline
\end{tabular*}
\end{center}
\end{table}

Before concluding, we comment on the relative contributions of the gluino,
chargino, and the rest (namely, neutralino, charged Higgs, and SM
contributions) to the total cross section. In Table~\ref{compare}, we
   show the relative contributions to $\sigma(pp\to
t\bar{c}+\bar{t}c+X)$ from gluino, chargino, and the rest, in the MSSM
with GUT mass relations (the case without GUT mass relations is shown for
$m_{\tilde{g}}=300$ GeV in brackets if different). For simplicity we
set  $(\delta_{U}^{23})_{AB}=(\delta_{D}^{23})_{AB},\, A,B=L,R$ and
the values $0,~ 0.2,$ and $0.4$ are considered. The other parameters
are $A=400$ GeV, $\tan\beta=10$, $M_{\rm {SUSY}}=\mu=250$ GeV and
$M_2=200$ GeV. The case in which no GUT relations between gaugino
masses are imposed corresponds (in our analysis) to the case in which
the gluino mass is allowed to be smaller. This is the reason why only the
gluino contributions  are enhanced in this scenario. The gluino
contributions are also dominant for the case in which GUT relations
are imposed, and the chargino contributions are two orders of
magnitude smaller.
However, one could envisage a case in which the
SUSY FCNC parameters $(\delta_{D}^{23})_{AB}$ are allowed to be
large, while the ones in the up sector restricted to be small or
zero, in which the chargino contribution could be
dominant. In the case of the constrained MSSM only chargino loops contribute.
\begin{figure}[htb]
\vspace{-0.1in}
\centerline{\hspace*{-0.9cm} \epsfxsize 3.5in
{\epsfbox{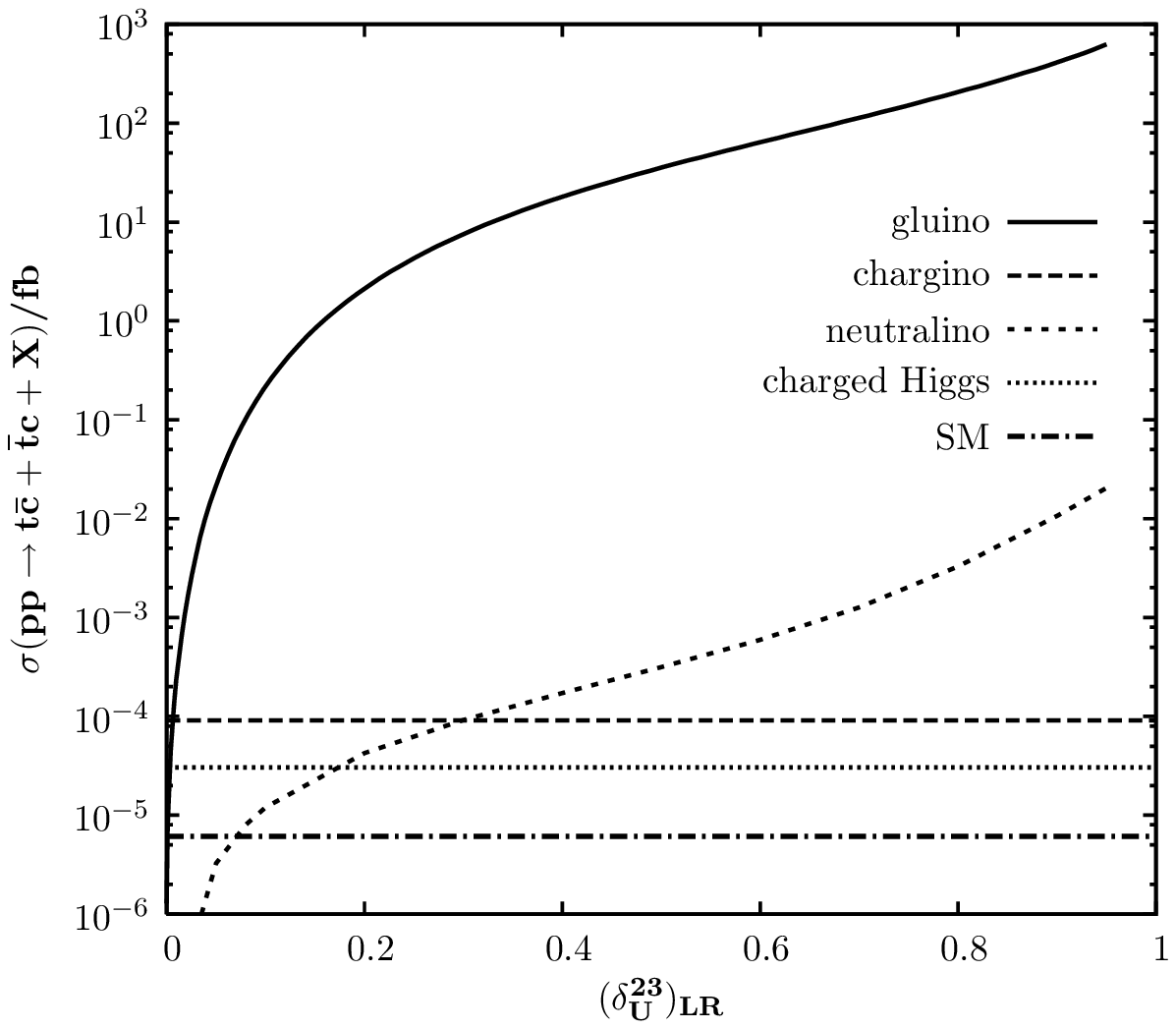}}}
\vskip -0.2in
\caption{\texttt{The total hadronic cross section
$\sigma(pp\to t\bar{c}+\bar{t}c+X)$ via gluon fusion as a function of
$(\delta_U^{23})_{LR}$ (other flavor violating parameters are assumed zero)
for $\tan\beta=30$, $m_{A^0}=500$ GeV,
$A=300$ GeV, $\mu=M_{\rm {SUSY}}=400$ GeV, $M_2=200$ GeV,
and $m_{\tilde{g}}=200$ GeV.}}\label{fig:best}
\end{figure}

   So far, we have considered a common flavor
violation in the LL, LR, RL, and RR sectors of sfermion matrix.
However, one could
analyze  the relative effects of flavor violation in each sector and determine
how large their contribution to the  cross section could be. If the
flavor violation comes only from the up sector,
then $(\delta_U^{23})_{LR}$ is the most sensitive parameter, as shown in
Fig.~\ref{fig:best}
   where we took the parameter values $m_{A^0}=500$ GeV, $\mu=M_{\rm
SUSY}=400$ GeV, $A=300$ GeV,
$m_{\tilde{g}}=200$ GeV, and $\tan\beta=30$. Of course this is only
one of the best possible scenarios. As seen from the figure, the
total cross section can be
as big as $630$ fb and the gluino contribution becomes dominant if
there exist a large flavor
violation in the up LR sector between the second and third
generations. If the flavor
violation comes only from the down sector, $(\delta_D^{23})_{LL}$
gives the largest contributions.
In this case the cross section is dominated by chargino contribution
and  can be as large as $0.4$ fb
with the same parameter values.

  Finally, we would like to qualitatively comment on the observability
of both the decay and the production channels of the top quark considered here.
The rare decay $t\to cgg$ can in general be treated twofold way: one can either
treat it inclusively  with $t\to cg$ or consider it as a separate channel.
The former means that $t\to cgg$ is taken as QCD-correction to $t\to cg$ by
  assuring that two of three final state jets are collinear so that
only two can be resolved in the detector. The latter can be a
competitive possibility if $BR(t\to cgg)$ is significantly larger than that
of $t\to cg$.  In here, and in our previous work \cite{EFTSM},
we conclude that $t\to cgg$ could be potentially more significant than the
two-body decay channel in the SM \cite{EFTSM} and in some
part of the MSSM  parameter space. However, this should be
taken with a dose of caution. Collinearity should be
avoided by applying certain cuts.  The unphysical C-parameter introduced in
the phase space  here plays an
essential role to distinguish $t\to cg$ from $t\to cgg$. Even though in our
explorations we considered several values of C, C must be taken
  in the range  of jet energy
resolution of the upcoming LHC detector. For $C\ll 0.1$,  $t\to cgg$
dominates over  $t\to cg$ for a larger parameter space, thus availability of
better jet resolution would give an opportunity to detect $t\to cgg$ before
$t\to cg$.

At LHC, predominantly $t\bar{t}$ pairs are produced. If we consider 
one of the top quarks
decay mainly as
  $t\to bW$ and the other one exotically as $t\to cgg$, then  the 
signal would be
$pp\to t\bar{t}\to (l\nu)ggc{\bar b}$ (4-jets, a lepton and missing energy), 
where $l=e,\mu$.\footnote
{Considering the W decay hadronically would produce a 6 jet final 
state, requiring
determination of the multi-jet trigger threshold.} For the single 
lepton plus jet topology,
it is possible to reconstruct the final state fully, and the b-quark can be
tagged to obtain  a cleaner signal, which introduces extra selection 
efficiency. We assume $\sigma(pp\to t{\bar t})=800$ pb at LHC and also that the W boson decays leptonically, not hadronically. Under this conditions, one can calculate roughly the total expected(raw) number of events as
$$N=\sigma(pp \to t{\bar t})\times BR({\bar t}\to{\bar b} W)\times BR(W\to l\nu)\times L\times BR(t\to cgg)\,,$$
where $L$ is the integrated luminosity which we take as 100 fb$^{-1}$. Therefore we have
$N = 800\times 10^3\times 1 \times (2/9)\times 100\times BR(t \to cgg)
  = (1.77\times10^7)\times BR(t\to cgg)$.   So, one expects around $(1.8\times 10^7)\times{BR}(t\to cgg)$ lepton+4 jets events
for an integrated
  luminosity $100 fb^{-1}$. However, counting a total efficiency including
trigger efficiency,  selection efficiency, as well as detector geometrical
acceptance, one would approximate a  total efficiency around 1\% \cite{brigitte}. Thus, the
number reduces to $(1.8\times 10^5)\times{BR}(t\to cgg)$.  So, if the flavor
violation comes from the down squark sector, then for  most part of the
parameter  space $t\to cgg$ would dominate over $t\to cg$ by around 
two orders of
magnitude, but both will remain unobserved because lepton+jets events are
less than a single event. If the flavor violation comes from up-squark sector, then
$t\to cg$ dominates and can reach $10^{-5}$ level for light gluino scenarios,
which might lead an observable  event rate around 1.8. Should the integrated luminosity increase at later runs of LHC, one could obtain larger event rates (up to 10 events) if the flavor violation is driven by the up squark sector. If the flavor violation comes from the down sector of the unconstrained MSSM, the event rate remains below the observable level.

For the single top production case $pp\to t\bar{c}+\bar{t}c+ X$, we
already included  cuts for the  transverse momentum and rapidity of the
charm and top quarks in the final state, as well as a lower cut for  jet
separation. In this case, if assume that the top quark is going to be
reconstructed in the final state  one can predict for example 50,000 events for
an integrated luminosity $\rm 100\,fb^{-1}$ and
$\sigma(pp\to t\bar{c}+X)=0.5$ pb. A similar total efficiency 
consideration will
going to  reduce this further but there exist enough events to find
a signal under the best-case scenario.  Anything beyond the above qualitative
discussion about the observability of the decay and production  channels will
be considered in more detail in our future paper \cite{ongoing}.

\section{Conclusion}\label{sec:conclusion}
In
this study we analyzed two related issues in top quark physics.
In the first part of the paper, we concentrated on the
comparison of two rare top quark decays, $t\to cgg$ versus $t\to cg$,
within the unconstrained MSSM,  driven by mixing between the second
and third generations only. To the best of our
knowledge, $t\to cgg$ decay has been considered only in our recent
study \cite{EFTSM} within the SM framework, where
$ BR(t\to cgg)$ was found to be two orders of magnitude larger than $
BR(t\to cg)$. However, in the SM,  $ BR(t\to cgg)$ remains at $10^{-9}$
level, and thus too small to be detectable. Any
experimental signature of such channel would require the existence of
physics beyond the SM which justifies further analyses.
Here we studied this decay in the MSSM framework by allowing
non-zero flavor off-diagonal
parameters. Our conclusion of the dominance of the branching ratio of
$t \to cgg$ over $t \to cg$ in the SM paper remains mostly valid in the
MSSM framework,
     but now the $ BR$'s can become as large as 
$10^{-6}-10^{-5}$.\footnote{The cutoff parameter C=0.1 is 
being used.}  For the
cases in which
we impose the GUT relation between gaugino masses {bf and assume a 
flavor violation
in down squark sector},  the {\large large}
difference in
ratio between the $t\to cgg$ and $t\to cg$ modes disappears only in the
case of very large intergenerational flavor-violating parameters
(close to 1, or to their maximally allowed upper values).  In that case, $t\to
cg$  exceeds $t\to cgg$. In the case of non-zero 
$(\delta_{U}^{23})_{AB}$, $t\to
cg$ dominates
  $t\to cgg$, except in regions of small flavor violation.
Once we relax the GUT constraints, there is no longer
such a large difference between the two and three body decays, as long
as a small flavor violation
is turned on.  Once the
flavor off-diagonal parameters, $(\delta_{U,D}^{23})_{AB}$, are
introduced, the difference in branching ratios disappears as the
parameters are minute and  $ BR(t\to cg)$ becomes around 5 times 
larger than $ BR(t\to cgg)$.
As expected, if the SUSY-GUT
relations hold, both modes cannot exceed $10^{-7}$ level (except for
  flavor violating parameters near their maximum allowed values for $t\to cg$
decay), because the gluino mass is large. Once we relax this condition, both $t
\to cgg$ and $t \to cg$  have branching ratios of the order $10^{-6}-10^{-5}$
and $10^{-5}$,  respectively.

Having shown that the three body rare decay $t\to cgg$ is
indeed important (comparable with, or larger than, the two body decay $t\to
cg$), we carried out a complete calculation of the single
top-charm associated production at LHC via gluon fusion at partonic
level within the same scenarios discussed above. This production
cross section has been considered before including only the SUSY-QCD
contributions \cite{Liu:2004bb}. We performed a complete analysis by
including all the
electro-weak contributions: the chargino-down-type
squark, neutralino-up-type squark, charged Higgs, as well as the SM
contributions. For simplification, a common SUSY FCNC parameter
$\delta$ is assumed, $(\delta_{U, D}^{23})_{LL}=(\delta_{U,
D}^{23})_{RR}=(\delta_{U, D}^{23})_{LR}=(\delta_{U, D}^{23})_{RL}$
(in the up and down squark sectors), but most often only one common $\delta$
parameter in either sector is allowed to be non-zero each  time.  We
have shown that,
in the most promising scenarios (if a common SUSY FCNC parameter
$\delta$ is assumed), the total hadronic cross section
$\sigma(pp\to t\bar{c}+\bar{t}c+X)$ can become as large as $50-60$ fb and
could reach a few hindered fb, especially if we relax the GUT
relations between the gaugino masses and assume a flavor violation
from one sector only
   (LL, LR, RL, or RR). We have  shown that the cross section could be
as large as $600-700$
   fb if  a large flavor violation coming from only
$(\delta_U^{23})_{LR}$ is allowed.

The comparative gluino,
chargino and other contributions to the process have been
estimated. The gluino contributions dominate over most of the
parameter space, when allowing flavor violation in both up and down
squark mass matrices to be of the same order of magnitude. However,
the chargino contribution is non-negligible and would be dominant in
either the constrained MSSM, or if the flavor violation was allowed
to be much larger in the down than in the up squark sector. While the
chargino loop is two orders of magnitude smaller than the gluino for
$(\delta_{U}^{23})_{AB}=(\delta_{ D}^{23})_{AB}$, the contribution of
the ``rest" (neutralino, charged Higgs and SM) is around half of the chargino
contribution, for all of the SUSY FCNC parameters
chosen.

Gluon fusion could be more promising than $cg\to t$, $q
{\bar q}\to t {\bar c}$, or $cg\to gt$, which we leave for a further
study  \cite{ongoing}.

The huge differences in prediction between the constrained
and the unconstrained MSSM scenarios  make LHC a fertile testing
ground for the study of SUSY FCNC processes.  Any significant rate
for the top-charm associated production would be a signal of physics
beyond the SM, and in particular, of new flavor physics.

\section{Acknowledgments}
The
work of M.F. was funded by NSERC of Canada (SAP0105354).
The work of G.E. was supported in part by the Israel Science Foundation
and by the Fund for the Promotion of Research at the Technion.
G.E would like to thank J.J. Cao for helpful discussions.
I.T. would like to thank Micheal Rauch for
his help and suggestions about the use of the \texttt{HadCalc}
program.

{\it Note added.} After submitting the first version of the present
paper to the active
(hep-ph/0601253.v1)  we became aware of an
analysis similar to ours \cite{Guasch:2006hf}. They analyze  the QCD SUSY
contribution,
and take as flavor
violation only $(\delta_U^{23})_{LL}\ne 0$, while we switched all
flavor violating
parameters on between the second and third generations. In our
analysis a non-zero
$(\delta_U^{23})_{LR}$ parameter is dominant (by one order of
magnitude over $(\delta_U^{23})_{LL}$), which makes a comparison of
our results to theirs difficult.

\end{document}